\documentclass{aa} 
\pdfoutput=1
\usepackage{txfonts}
\usepackage{longtable}  
\usepackage{rotating}
\usepackage{natbib}
\usepackage{graphicx}
\usepackage{graphics}
\usepackage{psfrag}
\usepackage{amssymb}
\bibpunct{(}{)}{;}{a}{}{,}
\def\Teff{$T_{\mathrm{eff}}$}
\def\logg{\ensuremath{\log g}}
\def\vmic{$\upsilon_{\mathrm{mic}}$}
\def\vmac{$\upsilon_{\mathrm{macro}}$}
\def\vsini{\ensuremath{{\upsilon}\sin i}}
\def\kms{$\mathrm{km\,s}^{-1}$}

\def\exc{$\chi_{\mathrm{excit}}$}
\def\loggf{log$gf$}
\def\vr{${\upsilon}_{\mathrm{r}}$}
\def\espa{ESPaDOnS}
\def\mus{MuSiCoS}
\def\nlte{non-LTE}
\def\llm{{\sc LLmodels}}
\def\atlas{{\sc ATLAS9}}
\def\tauros{\tau_{\rm ross}}
\begin{document} 
\title{Late stages of the evolution of A-type stars on the main sequence: 
comparison between observed chemical abundances and diffusion models for 8 Am 
stars of the Praesepe cluster} 
\subtitle{} 
\author{L. Fossati\inst{1}       \and 
	S. Bagnulo\inst{2}       \and 
	R. Monier\inst{3}        \and 
	S. A. Khan\inst{4}       \and 
	O. Kochukhov\inst{5}     \and 
	J. Landstreet\inst{4}    \and 
	G. Wade\inst{6}          \and 
	W. Weiss\inst{1}} 
\offprints{L.~Fossati} 
\institute{
	Institut fur Astronomie, Universit\"{a}t Wien, 
	T\"{u}rkenschanzstrasse 17, 1180 Wien, Austria.\\
	\email{fossati@astro.univie.ac.at; weiss@astro.univie.ac.at} 
	\and 
	Armagh Observatory, College Hill, Armagh BT61 9DG, Northern Ireland.
	\email{sba@arm.ac.uk} 
	\and 
	Groupe de Recherches en Astronomie et Astrophysique du Languedoc 
	UMR 5024, Universit\'{e} Montpellier II, Place E. Bataillon, 
	34095 Montpellier, France.
	\email{Richard.Monier@graal.univ-montp2.fr}
	\and 
	Department of Physics \& Astronomy, University of Western Ontario, 
	London, N6A 3K7, Ontario, Canada.\\
	\email{skhan@astro.uwo.ca; jlandstr@astro.uwo.ca}
	\and 
	Department of Astronomy and Space Physics, Uppsala University, 
	751 20, Uppsala, Sweden. 
	\email{oleg@astro.uu.se}
	\and  
	Physics Dept., Royal Military College of Canada, PO Box 17000, 
	Station Forces, K7K 4B4, Kingston, Canada.
	\email{Gregg.Wade@rmc.ca}
} 
\date{Received / Accepted} 
\abstract 
{}
{We aim to provide observational constraints on diffusion models that predict
peculiar chemical abundances in the atmospheres of Am stars. 
We also intend to check if chemical peculiarities and slow rotation can be 
explained by the presence of a weak magnetic field.} 
{We have obtained high resolution, high signal-to-noise ratio spectra of eight 
previously-classified Am stars, two normal A-type stars and one Blue 
Straggler, considered to be members of the Praesepe 
cluster. For all of these stars we have determined fundamental parameters 
and photospheric abundances for a large number of chemical elements, with a 
higher precision than was ever obtained before for this cluster. For seven of 
these stars we also obtained spectra in circular polarization and applied 
the LSD technique to constrain the longitudinal magnetic field.} 
{No magnetic field was detected in any of the analysed stars. HD~73666, a Blue
Straggler previously considered as an Ap\,(Si) star, turns out to have the 
abundances of a normal A-type star. Am classification is not confirmed for 
HD~72942. For HD~73709 we have also calculated synthetic $\Delta a$ 
photometry that is in good agreement with the 
observations. There is a generally good agreement between abundance 
predictions of diffusion models and values that we have obtained for the 
remaining Am stars. However, the observed Na and S abundances deviate from 
the predictions by 0.6 dex and $\geq$0.25 dex respectively. Li appears to be 
overabundant in three stars of our sample.} 
{} 
\titlerunning{Chemical abundance analysis of A-type stars in the Praesepe
cluster}
\authorrunning{L.~Fossati et al.}
\maketitle
\section{Introduction}\label{introduction}
Main sequence A-type stars present spectral peculiarities, usually interpreted 
as due to peculiar photospheric abundances and abundance distributions which 
are believed to be produced by the interaction of a large variety of physical 
processes (e.g. diffusion, magnetic field, pulsation and various kinds of 
mixing processes).

An interesting problem that has yet to be addressed is how these 
peculiarities change during main sequence evolution. The chemical composition 
of field A-type stars have been studied by several authors, e.g. 
\citet{hillland1993}, \citet{adelman2000}. However, it is not straightforward 
to use the results of these investigations to study how photospheric chemistry 
evolves during a star's main sequence life. First, the original composition 
of the cloud from which stars were born is not known and is likely somewhat 
different for each field star. It is therefore not possible to discriminate 
between evolutionary effects and differences due to original chemical 
composition. Secondly, it is difficult to estimate the age of field stars 
with the precision necessary for such evolutionary studies 
\citep[for a discussion of this problem see][]{stefano2006}.

From this point of view, A-type stars belonging to open clusters are much 
more interesting objects. Compared to field stars, A-type stars in open 
clusters have three very interesting properties:
\begin{list}{-}{}
 \item they were all presumably born from the interstellar gas with an 
 approximately uniform composition;
 \item they all have approximately the same age (to within a few Myr);
 \item their age can be determined much more precisely than for field stars.
\end{list}

Few abundance analyses of A-type stars in open clusters have been carried 
out. Those that have been published have usually focused on a limited numbers 
of stars.

\citet{monier1999} have determined the abundances of eleven chemical elements 
for a large sample of stars regularly distributed in spectral type along the 
main sequence in order to sample the expected masses uniformly.  All these 
stars were analysed in a uniform manner using spectrum synthesis. 
\citet{ch2006} have performed a detailed abundance analysis for five A-type 
stars of the young open cluster IC~2391.
\citet{folsom2007} have performed a detailed abundance analysis for four 
Ap/Bp stars and one normal late B-type star of the open cluster NGC~6475.
 
A goal of this programme is to determine photospheric abundance patterns in 
A-type star members of clusters of different ages. This is crucial in order to: 
\textit{i)} investigate the chemical differences between 
normal and peculiar stars inside the same cluster, \textit{ii)} study the 
evolution with time of abundance peculiarities by studying clusters of 
various ages, \textit{iii)} set constraints on the hydrodynamical processes 
occurring at the base of the convection zone in the non magnetic stars and 
\textit{iv)} study the effects of diffusion in the presence of a magnetic 
field for the magnetic (Ap) stars in the cluster. The abundance analysis will 
be performed in an homogeneous way applying a method described in this first 
work.

Praesepe (NGC~2632), a nearby intermediate-age open cluster
\citep[$\log t = 8.85\pm 0.15$,][]{gonzalez},
is an especially interesting target because it includes a large number of 
A-type stars, among which are many Am stars. Furthermore, because the 
cluster is relatively close to the sun 
\citep[d~=~180~$\pm$~10 pc,][]{robichon1999}, many of the member A-type stars 
are bright enough to allow us to obtain high resolution spectra 
with intermediate class telescopes. 

We dedicate this first paper to the Am stars of the Praesepe cluster,
searching for magnetic fields in these objects and discussing 
the differences between "normal"\footnote{We consider as "normal" A-type stars 
all the A-type stars that are classified neither as Am nor Ap} A-type stars 
and Am stars in the cluster. 

We also compare our results with previous works and with theoretical 
chemical evolution models. In particular we take into account diffusion models 
by \citet{richer2000}. We want to provide observational constraints to
the theory of the evolution of the abundances in normal and chemically peculiar
stars. Our detailed abundance analysis could provide information about the 
turbulence occurring in the outer stellar regions in Am stars with well
determined age. In particular, our analysis can give constraints to define 
the depth of the zone mixed by turbulence, since it is the only parameter 
characterising turbulence \citep{richer2000}. A systematic abundance analysis 
of normal and peculiar stars in clusters could provide information on the 
origin of the mixing process and show if only turbulence is needed to explain 
abundance anomalies, or if other hydrodynamical processes occur.

We tackle this problem using new and more precise new-generation spectrographs 
providing a wider wavelength coverage together with newer analysis codes and 
procedures (e.g. Least-Squares Deconvolution and synthetic line profile 
fitting instead of equivalent width measurements.)

The observed stars, the instruments employed and the target selection are 
described in Sect.~\ref{observations}. The data reduction and a discussion of
the continuum normalisation are provided in Sect.~\ref{reduction,norm}. In 
Sect.~\ref{LLmodels} and \ref{abundanceanalysis} we describe the models and
the procedure used to perform the abundance analysis. Our results are 
summarised in Sect.~\ref{results}. Discussion and conclusions are given in 
Sect.~\ref{discussion} and \ref{conclusion} respectively.
\section{Observations and data reduction}\label{observations}
\subsection{Instruments}
We observed six stars of the Praesepe cluster using the \espa\, (Echelle 
SpectroPolarimetric Device for Observations of Stars) spectropolarimeter 
at the Canada-France-Hawaii Telescope (CFHT) from January 8th to 10th 2006. 
Spectra were acquired in circular polarisation.

Spectra of an additional four stars were obtained with the ELODIE spectrograph 
at the Observatoire de Haute Provence (OHP) from January 4th to 6th 2004.

An additional circular polarisation spectrum of HD~73709 was obtained with 
the \mus\, spectropolarimeter at the 2-m Bernard Lyot Telescope (TBL) of the 
Pic du Midi Observatory on 7 March 2000. The magnetic field derived form this 
observation has been discussed by \citet{shorlin2002}.
\subsubsection{\espa}
\espa\, consists of a table-top cross-dispersed echelle spectrograph fed via 
a double optical fiber directly from a Cassegrain-mounted 
polarization analysis module. In "polarimetric" mode, the instrument can 
acquire a Stokes $V, Q$ or $U$ stellar spectrum throughout the spectral 
range 3700 to 10400 \AA\, with a resolving power of about 65\,000. A complete 
polarimetric observation consists of a sequence of 4 sub-exposures, between 
which the retarder is rotated by $\pm$ 90 degrees
\citep{donatietal1997,wade2000}. 
In addition to the stellar exposures, a single bias spectrum and ThAr 
wavelength calibration spectrum, as well as a series of flat-field 
exposures, were obtained at the beginning and end of each night.
\subsubsection{ELODIE}
ELODIE is a cross-dispersed echelle spectrograph at the 1.93-m 
telescope at the OHP observatory. Light from the Cassegrain focus is fed 
into the spectrograph through a pair of optical fibers. Two focal-plane 
apertures are available (both 2 arc-sec wide), one of which is used for 
starlight and the other can be used for either the sky background or the 
wavelength calibration lamp, but can also be masked. The spectra cover a 
3000 \AA\, wavelength range (3850--6800 \AA) with a mean spectral resolution of 
42\,000. ELODIE was decommissioned in mid-August 2006.
\subsubsection{\mus}
\mus, like \espa, consists of a table-top cross-dispersed echelle spectrograph
fed via a double optical fiber directly from a Cassegrain-mounted polarization
analysis module. \mus\, provides a Stokes $V$, $Q$ or $U$
stellar spectrum from 4500 to 6600 \AA\, with a
mean resolving power of 35\,000. A more detailed description of the
instrument and of the observing procedures are reported by 
\citet{donatietal1999}. \mus was decomissioned in December 2006.
\subsection{Target selection}
The target selection for the run with the \espa\ spectrograph was performed 
taking into account previously-published peculiar spectral classifications of 
the stars, together with their \vsini\ (if any), giving priority 
to slowly-rotating stars. Data concerning stars of the cluster were collected 
from the WEBDA database\footnote{{\tt http://www.univie.ac.at/webda}} 
\citep{webda} and the SIMBAD database, operated at CDS, Strasbourg, France.

The stars observed with the ELODIE spectrograph were the bright stars included 
in the analysis provided by \citet{burkcoupry1998}, to allow a comparison 
with their work.

The star observed with \mus\ was analysed by \citet{shorlin2002},
with the LSD technique, to measure the magnetic field strength.

Four stars of the sample were accepted as cluster members by the HIPPARCOS
survey \citep{robichon1999} whereas the others have been confirmed as members 
by different studies, such as those by \citet{kharchenko2004} 
and \citet{wang1995}.

The complete sample of stars observed and analysed in this paper is listed 
in Table \ref{tabella radec}. Seven of the stars are spectroscopic 
binaries and one is a $\delta$ Scuti star. Of the eleven stars observed, eight 
were previously classified as Am stars, two as normal A-type stars and one 
as an Ap\,(Si) star.  

\begin{table*}[ht]
\caption[ ]{Basic datas of the observations for the program stars. $^{1}$: positions from 
\citet{perrymanetal1997}; $^{2}$: positions from \citet{hogetal1998}. The SNR are 
calculated at $\sim$5000~\AA\ in a bin of 0.5 \AA. The exposure time is in seconds. With the 
\espa\, and \mus\, instruments we obtained Stokes $I$ and $V$ spectra, which allowed us to 
attempt magnetic field detection. With ELODIE we obtained Stokes $I$ spectra only.
The HJD indicate the Heliocentric Julian Date at the middle of the exposure.}
\label{tabella radec}
\centering                      
\scriptsize{
\begin{tabular}{cccccccccccl}
\hline
\hline
HD & RA & DEC & HJD & M$_{\it{v}}$ & Spectral Type & Instrument & Resolution & SNR & Exp. Time & Remarks \\
\hline
73430 & 08 39 03.585  & $+$19 59 59.08~$^{2}$  & 2453746.125 & 8.33 & A9V   & \espa  & 65\,000 & 220 & 1800 &			 \\   
73575 & 08 39 42.6548 & $+$19 46 42.440~$^{1}$ & 2453747.116 & 6.66 & F0III & \espa  & 65\,000 & 250 & 2400 & $\delta$Sct variable \\	
73666 & 08 40 11.4528 & $+$19 58 16.073~$^{1}$ & 2453745.063 & 6.61 & Ap(Si)& \espa  & 65\,000 & 660 & 1600 & SB1, Blue Straggler  \\	
72942 & 08 36 17.4422 & $+$20 20 29.421~$^{1}$ & 2453746.098 & 7.48 & Am    & \espa  & 65\,000 & 350 & 1600 & SB?		 \\   
73045 & 08 36 48.0033 & $+$18 52 58.111~$^{1}$ & 2453745.091 & 6.82 & Am    & \espa  & 65\,000 & 290 & 2400 & SB1		 \\   
73730 & 08 40 23.466  & $+$19 50 05.91~$^{2}$  & 2453745.123 & 7.99 & Am    & \espa  & 65\,000 & 330 & 2000 &			 \\   
73618 & 08 39 56.496  & $+$19 33 10.76~$^{2}$  & 2453009.612 & 7.30 & Am    & ELODIE & 42\,000 & 160 & 3600 & SB1		 \\   
73174 & 08 37 36.995  & $+$19 43 58.48~$^{2}$  & 2453010.574 & 7.76 & Am    & ELODIE & 42\,000 & 120 & 2700 & SB1, triple system   \\	
73711 & 08 40 18.099  & $+$19 31 55.17~$^{2}$  & 2453010.432 & 7.51 & Am    & ELODIE & 42\,000 & 90  & 3600 & SB1		 \\   
73818 & 08 40 56.935  & $+$19 56 05.47~$^{2}$  & 2453012.102 & 8.69 & Am    & ELODIE & 42\,000 & 80  & 2400 & SB1		 \\   
73709 & 08 40 20.748  & $+$19 41 12.24~$^{2}$  & 2451611.473 & 7.68 & Am    & \mus   & 35\,000 & 120 & 2400 & SB1, quadruple system\\	
\hline
\end{tabular}
}
\end{table*}



%
\subsection{Data reduction}\label{reduction,norm}
The \espa\ spectra were reduced using the Libre-ESpRIT package 
\footnote{\tt www.ast.obs-mip.fr/projets/espadons/espadons.html -  
See also \citet{donatietal1997}}.

The ELODIE spectra were automatically reduced by a standard data reduction 
pipeline described by \citet{baranneetal1996}.

The \mus\, spectrum was reduced according to the procedure described by 
\citet{wade2000} and \citet{shorlin2002}.

The sample of stars includes objects with a high \vsini\ (up to 
$\simeq 130$\,\kms) for which the continuum normalisation is a critical 
reduction procedure. For this reason, all of the spectra were normalised 
without the use of any automatic continuum fitting procedure.
We considered the single echelle orders of the spectra, which were normalised 
and then merged. It was not possible to determine a correct continuum level 
short wards of the H$\gamma$ line (4340.462 \AA), as there were not
enough continuum windows in the spectrum at these shorten wavelengths. 
\section{Calculation of model atmospheres}\label{LLmodels}
Model atmospheres were calculated with the LTE code \llm\ (version 8.4), which 
uses direct sampling of the line opacity \citep{denis2004}, and allows the 
computation of model atmospheres with individualised (not scaled solar) 
abundance patterns. This allows us to compute self-consistent model 
atmospheres that match the actual abundances of chemically peculiar stars and 
to thereby minimise systematic errors \citep{sergey2007}.

We used the VALD database \citep{vald1,vald2,vald3} as a source of spectral 
line data, including lines that originate from predicted levels. We then 
performed a preselection procedure to eliminate those lines that do not 
contribute significantly to the line opacity. For this procedure we utilised 
model atmospheres calculated by \atlas\ \citep{kurucz1993a} with fundamental 
parameters corresponding to each star in our sample. Because at this stage of 
the analysis we did not know the photospheric abundances of the sample stars, 
we employed the Opacity Distribution Function (ODF) tables for solar 
abundances \citep{kurucz1993b}. The line selection criterion required that 
the line-to-continuum opacity ratio at the center of each line, at any 
atmospheric depth, be greater than 0.05~\%.

To compute atmosphere models with individual abundance patterns, we used an 
iterative procedure. The initial model atmosphere was calculated with the 
solar abundances taken from \citet{asplundetal2005}. Then, these abundances 
were modified according to the results of spectroscopic analysis, and a final 
model atmosphere was iterated.

A logarithmic Rosseland optical depth scale ${\log\tauros}$ was adopted as 
an independent variable of atmospheric depth spanning from $+2$ to $-6.875$ 
and subdivided into 72 layers. Opacities were sampled with a 0.1\,\AA\ 
wavelength step. Since {\it a posteriori} we found no magnetic field in the 
atmospheres of our stars (Sect.~\ref{am as magnetic}), the excess line 
blanketing due to a magnetic field \citep{oleg2005,llm2,llm3} was neglected. 
The value of the microturbulence velocity \vmic\, was adopted according to 
the results of the spectroscopic analysis (Sect.~\ref{howparameters}).

Convection was treated according to the CM approach \citep{CM}. 
It has been argued that CM should be preferred over the MLT approach 
\citep{FrizK}. For example, \citet{SK} showed that the CM models give results 
that are generally superior to standard MLT (${\alpha=1.25}$) models. 
Furthermore, to calculate a part of the NEMO model atmosphere grid, 
\citet{heiter} adopted the free parameter ${\alpha=0.5}$ for the MLT approach 
as it produced results quite similar to those of the CM method. Consequently, 
we have taken convection (CM approach) into account for the whole set of 
calculations to ensure correct modelling for lower effective temperatures, 
as many of the stars analysed in this paper have effective temperatures 
below 9000\,K. To test the importance of the convection treatment on the 
abundance analysis, as applied to this sample of stars, we performed several 
numerical tests using model atmospheres calculated with no convection 
treatment at all. The results revealed differences of about 0.005\,dex, 
clearly within the errors associated with the analysis.
\section{Spectral analysis}\label{abundanceanalysis}
\subsection{High precision search for magnetic field}
One of the main goals of our analysis is to search for weak magnetic fields in 
the Am stars of Praesepe, and to check if the Ap\,(Si) star HD~73666 has one, 
since many Ap stars show a strong magnetic field. 
For these reasons, we observed some Am stars and HD~73666 with the \espa\ 
spectrograph that provides the opportunity to obtain high resolution spectra in
circular polarization. To detect the presence of a magnetic field and to infer 
the longitudinal magnetic field we used the Least-Squares Deconvolution 
technique (hereafter LSD). 
LSD is a cross-correlation technique developed for the 
detection and measurement of weak polarization signatures in stellar 
spectral lines. The method is described in detail by 
\citet{donatietal1997} and \citet{wade2000}.
%
%
\subsection{Atmosphere fundamental parameters}\label{howparameters}
The initial values of the effective temperature (\Teff), surface gravity 
(\logg) and metallicity of our sample of stars were estimated via 
Str\"{o}mgren photometry \citep{hauck}. 

The initial value of the microturbulence velocity (\vmic) was determined 
using the following relation \citep{pace2006}:
\begin{equation}
\label{vmicphotometry}
\upsilon_{\mathrm{mic}}=-4.7\log(T_{\mathrm{eff}})+20.9\,\,\mathrm{km\,s}^{-1}\ .
\end{equation}
The uncertainties associated with \Teff, \logg\ and \vmic\ determined 
photometrically and from Eq. (\ref{vmicphotometry}) are quite large 
(about 300\,K in \Teff\ and 0.25 in \logg, due to uncertainties in the 
photometric measurements and intrinsic scatter in the calibrations). We have 
adjusted these parameters spectroscopically to get more accurate values for an 
abundance analysis.

The best value of the effective temperature was estimated using the 
abundance--\exc\ correlation, calculated fitting the abundances of different 
selected lines of an element, in the abundance--excitation potential plane. 
This correlation is sensitive to effective temperature variations. This 
property allows us to determine the best value of the effective temperature, 
which we derived by eliminating the abundance--\exc\ correlation. For this 
step, the abundances were calculated using a modified version \citep{vadim} 
of the WIDTH9 code \citep{kurucz1993a}, using equivalenth widths for the more 
slowly-rotating stars (\vsini\, $<$ 30 \kms), and by fitting line cores 
(as described in \ref{linecorefitting}) for the rapid rotators.
 
In a similar way, the best value for \logg\ was found by eliminating any 
systematic difference in abundance derived from different ionisation stages 
of the same chemical element. 

Finally, for stars with \vsini\ less than 30~\kms, the \vmic\, value was 
determined by eliminating any abundance--equivalent width correlation. 

To check the degree of refinement reached with the spectroscopic method, we 
systematically performed the following check. For each star with \vsini\,
$<$ 30 \kms\, we have performed the abundance analsysis of \ion{Fe}{i} and
\ion{Fe}{ii} lines using three different approaches, i.e., by calculating
\logg\, and \Teff\, from (i) Geneva photometry, (ii) from Str\"{o}mgren
photometry, and (iii) with the spectroscopic method outlined above. For each 
of these three models, we have calculated the abundances of \ion{Fe}{i} and
\ion{Fe}{ii} lines for \vmic\, = 0, 1, 2, ..., 6 \kms. Finally, for each of 
these 18 sets of \Teff, \logg\, and \vmic\, values, we have calculated the 
standard deviation from the mean \ion{Fe}{i} and \ion{Fe}{ii} values. In all
cases we found that the parameters that we have identified with the
method described above are those that minimise the abundance scatter.

\begin{figure}[ht]
\begin{center}
\resizebox{\hsize}{!}{\includegraphics{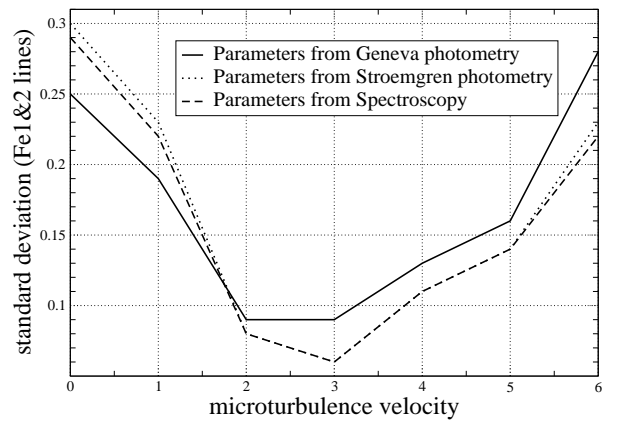}}
\caption{Standard deviation for \ion{Fe}{i} and \ion{Fe}{ii} lines, selected 
for HD~73730, as a function of \vmic\, using fundamental parameters obtained 
from Geneva photometry (straight line), from Str\"{o}mgren photometry (dotted 
line) and from spectroscopy (dashed line).}
\label{vmic}
\end{center}
\end{figure}

In Fig.~\ref{vmic} we give an example of the result of our check, for the Am
star HD~73730. The result of the check for all the other stars of the sample 
is similar to the one shown in Fig.~\ref{vmic}.
                                                  
For faster rotators an insufficient number of unblended lines are available 
with which to calculate the equivalent width. The more slowly-rotating Am 
stars of the sample show a mean \vmic\, value around 2.7 \kms. For this 
reason, for fast rotators, we retained the value found using 
Eq.~(\ref{vmicphotometry}), which was always compatible, within the errors, 
with 2.7 \kms.

Following \citet{john1998} it has been generally adopted that for Am 
stars \vmic\, is between 4--5 \kms. By contrast in the present work we 
determined values between 2.3--3.1 \kms, for the slowly rotating stars.
 
The chemical elements used to determine the fundamental parameters were 
different for each star, depending mainly on the value of \vsini, and are 
described in Sect.~\ref{results}.

For the lines used to eliminate the correlations discussed above we assumed 
that \nlte\ and stratification effects were negligible \citep{tanyasub}.

The wavelength range of the \espa\, orders was wide enough to perform a 
safe continuum normalisation of the H$\alpha$ and H$\beta$ lines. For 
the spectra obtained with this instrument it was therefore possible to use 
these two hydrogen lines to provide an additional constraint on the 
fundamental parameters. In particular, for effective temperatures less than 
8000\,K, the wings of the Hydrogen lines are very sensitive to the effective 
temperature. On the other hand, at higher effective temperatures the wings are 
particularly sensitive to \logg. We did not use the hydrogen lines of the 
spectra obtained with ELODIE and \mus\ because the wavelength ranges of the 
orders containing the H$\alpha$ and H$\beta$ lines were too narrow to include 
the entire line profile.

The spectroscopic procedure of refining the fundamental parameters reduces the 
errors of \Teff\ and \logg\ to typically 200\,K and 0.2, respectively, while 
the estimated error for \vmic\ is around 0.2\,\kms. These errors are estimated 
taking into account different noise sources, including continuum normalisation 
errors and uncertainty in the convective model. The errors associated with the 
effective temperature and \logg\, have been confirmed by a fit of the 
H$\alpha$ and H$\beta$ line profiles using the models calculated with the
parameters obtained from the estimated errors. 

In some cases a star's high rotational velocity prevented us from recovering 
\Teff\, and \logg\, spectroscopically with good precision. In these cases we 
ultimately adopted an average of parameters determined spectroscopically and 
those obtained via Str\"{o}mgren photometry. 

For the spectra obtained with the ELODIE spectrograph we also attempted to 
check for the presence of a magnetic field by looking for a correlation 
between the abundance derived from each line and its Land\'e factor. We also 
conducted a search for magnetically-split lines (e.g. \ion{Fe}{ii} line at
6149.258~\AA) in spectra of the very slowest rotators. In fact, for \vsini\, 
more than a few \kms\, this method cannot be used for detecting typical 
Ap-star magnetic fields. For the spectra obtained with the \espa\, 
spectrograph we used the LSD technique to measure the velocity-resolved 
Stokes $V$ profile and the longitudinal magnetic field.

With the correct fundamental parameters, we determined the best value
of \vsini\, fitting it on the \ion{Fe}{i} lines at 5434.524~\AA\, and 
5576.089~\AA. We chose these two \ion{Fe}{i} lines because their broadening 
parameters are well known and are not affected by magnetic field 
broadening, since their Land\'{e} factor is close to $0$. The estimated 
error on \vsini\, is about 5\%.

The radial velocity, in \kms, was determined by performing the median of 
the results obtained by fitting several lines of the observed spectrum to a 
synthetic one. 
\subsection{Element abundance analysis}\label{linecorefitting}
Once we had obtained the best values for the fundamental parameters, we 
determined the final elemental abundances by direct fitting of the observed 
spectra with synthetic models. We synthesised the model spectra with Synth3 
\citep{synth3} and fit the cores of selected lines to get a value of the 
abundance associated with each line. We then calculated the mean and the 
relative standard deviation for each analysed element. The line core fitting 
was performed with the 'Lispan' and 'ATC' codes (written by Ch. St\"{u}tz) 
together with Synth3. The only free parameter in the line core fitting 
procedure is the abundance of the line. The fitting procedure and the 
determination of the abundances was performed iteratively in order to obtained 
a better determination of the abundances for blended lines.
The error associated with the derived abundance of each element is the 
standard deviation from the mean abundance of the selected lines of that 
element. These errors do not take into account the uncertainties of the
fundamental parameters and of the adopted model atmosphere. To have an idea of
these uncerteinties, we have calculated the abundances of Fe, Ti and Ni for 
HD~73730 with five different models varing \Teff\, of $\pm$200 K and \logg\, of
$\pm$0.2 dex from the adopted model. We found a variation of less than 0.2 dex 
for Fe, 0.1 dex for Ti and 0.1 for Ni due to the temperature varation. We did 
not find any significant abundance change varing \logg. The uncertainty in
temperature is probably the main error source on our abundances.

The lines used to synthesise the spectra and selected to calculate the
abundances for the various elements were extracted from the VALD database.

The number of selected lines depends on the \vsini\, value found for each star.
We also checked the \loggf\, value of each selected line using the solar 
spectrum. Lines in the synthetic spectrum of the Sun showing large deviations 
from the observed solar spectrum were rejected from the selection. For some 
elements (e.g. Co, Cu and Zn) we derived the abundance from one line. This does 
not mean that only one line is present in the spectrum, but that only one line 
was selected to derive the chemical abundance. Other lines of the same element 
are present, but they were not selected because they are either too much 
blended or their \loggf\, value appeared to be in error based on the solar 
spectrum.  

For the elements which were not analysed we assumed solar abundances from 
\citet{asplundetal2005}.

%
\subsection{Non-LTE effects}
Some potassium lines were visible in the \espa\ spectra (because of the 
extension of the spectra into the far red), making possible the calculation 
of the abundance of this element. In particular the \ion{K}{i} line at 
7698.974\,\AA\ was detected and was largely unblended in the spectra of most 
stars. Unfortunately, this element presents strong \nlte\ effects; for this 
reason the K abundances derived here are reported only as upper limits. The 
He, O and Na lines were selected in such a way to reduce as much as possible 
\nlte\, effects. For He, we selected lines in the blue region; for O, lines in 
the red region; and for Na, we used the \ion{Na}{i} doublet at 
$5682.633 - 5688.205$\,\AA, which is known to be essentially uneffected by 
\nlte\, effects (Ryabchikova, private communication).
For all the other elements that present important \nlte\ 
effects (e.g. C, N, Al, S, Y, Ba), the abundances derived here should be 
considered as upper limits 
\citep[][and references therein]{baumueller,kamp2001,przybilla2006}. 
However at present day it is not known how large are the deviations from 
LTE for normal A-type stars and Am stars for the various elements. More \nlte\, 
analysis should be performed for this type of star. Since our sample of 
stars has similar fundamental parameters (mainly \logg\, and 
metallicity) the \nlte\, effects associated to each star should be of 
the same magnitude. This allows us to compare the stars of our sample 
with each other.  
\section{Results}\label{results}
We have organised our sample of program stars according to their literature 
classification as normal A-type stars (HD~73430, HD~73575), Blue Stragglers 
(HD~73666 -- Ap(Si) star) and Am stars (HD~72942, HD~73045, HD~73730, 
HD~73618, HD~73174, HD~73711, HD~73818, HD~73709). 

Table~\ref{tabella parametri fondamentali sample} lists the initial and 
final set of fundamental parameters obtained for each program star.
\begin{table*}[ht]
\caption[ ]{Initial and final atmospheric fundamental parameters for the 
analysed stars of the Praesepe cluster. The initial fundamental parameters 
are derived from photometry, the final with spectroscopy. The errors on 
the fundamental parameters are estimated to be 200\,K, 0.2\,\kms and 0.2 for \Teff,
\vmic\, and \logg\, respectively. The estimated error on \vsini\, is about 5\%.
The \vr\, is given in \kms. $\langle B_z\rangle$ indicate the Longitudinal Magnetic 
Field (measured in Gauss) calculated with the LSD tecnique. BS: Blue Straggler.}
\label{tabella parametri fondamentali sample}
\centering                      
\begin{tabular}{cccccccccccc}
\hline
\hline
 & \multicolumn{3}{c}{initial set} & \multicolumn{3}{c}{final set} & & & & & \\
\hline
HD & \Teff & \logg & \vmic  & \Teff & \logg & \vmic  & \vmac  & \vsini & \vr	& $\langle B_z\rangle$ & comments \\
   &   [K] & [cgs] & [\kms] &	[K] & [cgs] & [\kms] & [\kms] & [\kms] & [\kms]& G		       &	  \\
\hline
73430 & 7790 & 3.95 & 2.6 & 7660 & 4.02 & 2.6 & 0  & 73   & 33.7 & 1	$\pm$ 45  & normal \\
73575 & 7435 & 3.46 & 2.7 & 7300 & 2.92 & 2.7 & 0  & 127  & 33.4 & -215 $\pm$ 149 & normal \\
73666 & 9429 & 3.84 & 2.2 & 9382 & 3.78 & 1.9 & 0  & 10   & 34.1 & 6    $\pm$ 5   & SB1; BS; close to normal \\ 
72942 & 8237 & 3.80 & 2.5 & 8450 & 3.90 & 2.4 & 0  & 73   & 39.9 & 12	$\pm$ 31  & SB?; between normal \& Am \\
73045 & 7470 & 4.09 & 2.7 & 7570 & 4.05 & 3.6 & 10 & 10   & 27.9 & -1	$\pm$ 4   & SB1; Am \\
73730 & 8009 & 3.83 & 2.5 & 8070 & 3.97 & 2.6 & 0  & 29   & 36.6 & -12  $\pm$ 9   & Am \\
73618 & 8091 & 3.81 & 2.5 & 8170 & 4.00 & 2.5 & 0  & 47   & 15.1 & $-$            & SB1; Am \\
73174 & 7600 & 4.20 & 2.0 & 8350 & 4.15 & 2.9 & 0  & $<$5 & 2.9  & $-$            & SB1; Am \\
73711 & 8269 & 3.95 & 2.5 & 8020 & 3.69 & 2.5 & 0  & 62   & 23.4 & $-$            & SB1; Am \\
73818 & 7232 & 3.82 & 2.8 & 7232 & 3.82 & 2.8 & 0  & 66   & 22.5 & $-$            & SB1; Am \\
73709 & 8063 & 3.79 & 2.5 & 8070 & 3.78 & 2.3 & 0  & 10   & 26.7 & $-$            & SB1; Am \\ 
\hline
\end{tabular}
\end{table*}


The derived abundances are illustrated in 
Fig.~\ref{abbondanze per hd73430} to Fig.~\ref{abbondanze per hd73709}, and 
reported in Table~\ref{abbondanze del sample}.
\begin{table*}[ht]
\caption[ ]{Abundances ($\log(N_{X}/N_{tot})$) of the program stars with the 
estimated internal errors in units of 0.01 dex, in parenthesis. For comparison, 
the solar abundances \citep{asplundetal2005} are given in the 
last column. At.N. gives the Atomic Numbers of the elements. Abundances obtained 
from just one line have no error ($-$). Upper limits are denoted by $<$.}
\label{abbondanze del sample}
\rotatebox{90}{
\scriptsize{
\begin{tabular}{cccccccccccccc}
\hline
\hline
\multicolumn{2}{c}{ } & \multicolumn{2}{c}{"Normal" A-type stars} & \multicolumn{1}{c}{Blue Straggler} & \multicolumn{8}{c}{Am stars} & Solar \\
At.N.&Element& HD~73430 & HD~73575   & HD~73666    & HD~72942  & HD~73045    & HD~73730   & HD~73618    & HD~73174   & HD~73711   & HD~73818   & HD~73709   & Abundances \\
\hline
2 & He      &           &            &$-1.20(05)  $&	       &             &	          &	        &	     &	          &	       &	    &$-1.11 $\\ 
3 & Li      &$<-8.17(-)$&$<-8.50(-) $&             &$<-8.49(-)$&$-8.86(02)$  &$-8.74(-)  $&$<-8.58(-)  $&$-8.51(02) $&$<-8.43(-) $&$<-8.63(-) $&	    &$-10.99$\\  
6 & C       &$<-3.74(-)$&$<-3.61(-) $&$<-3.25(03) $&$<-3.85(-)$&$<-4.37(04) $&$<-4.32(47)$&$<-4.16(16) $&$<-4.08(06)$&$<-3.99(15)$&$<-3.38(07)$&$<-3.71(26)$&$-3.65 $\\  
7 & N       &$<-3.62(16)$&$<-3.93(-)$&$<-3.67(10) $&$<-3.84(-)$&$<-4.49(-)  $&$<-4.63(-) $&	        &$<-4.50(12)$&	          &	       &	    &$-4.26 $\\  
8 & O       &$-3.26(-) $&$-3.37(-)  $&$-3.06(07)  $&$-3.58(-) $&$-3.90(10)  $&$-3.96(09) $&$-3.65(04)  $&$-3.72(01) $&$-3.60(04) $&$-3.76(-)  $&$-3.86(08) $&$-3.38 $\\  
11& Na      &$-5.69(06)$&            &$-5.47(01)  $&	       &$-5.32(03)  $&$-5.36(01) $&$-5.34(03)  $&$-5.13(04) $&	          &$-5.35(01) $&$-5.34(06) $&$-5.87 $\\  
12& Mg      &$-4.79(15)$&$-4.63(-)  $&$-4.25(14)  $&$-4.51(03)$&$-4.52(04)  $&$-4.43(06) $&$-4.29(-)   $&$-4.21(05) $&$-4.63(05) $&$-4.51(16) $&$-4.56(06) $&$-4.51 $\\  
13& Al      &           &            &$<-5.34(06) $&	       &$<-5.41(01) $&$<-5.37(-) $&	        &$<-5.06(-) $&	          &	       &$<-4.91(16)$&$-5.67 $\\  
14& Si      &$-4.44(04)$&$-4.41(-)  $&$-4.30(01)  $&$-4.67(-) $&$-4.21(05)  $&$-4.21(05) $&$-4.51(-)   $&$-4.08(11) $&$-4.19(05) $&$-3.44(32) $&$-4.10(06) $&$-4.53 $\\  
15& P       &           &            &             &	       &	     &            &	        &$-5.23(-)  $&	          &	       &$-5.12(-)  $&$-6.68 $\\  
16& S       &$<-4.59(-) $&           &$<-4.47(-)  $&$<-4.48(-)$&$<-4.67(09) $&$<-4.54(03) $&$<-4.52(13)$&$<-4.37(03)$&$<-4.62(-) $&$<-4.36(26)$&$<-4.44(01)$&$-4.90 $\\  
18& Ar      &           &            &$-5.10(-)   $&	       &	     &            &	        &	     &	          &	       &	    &$-5.86 $\\  
19& K       &$<-6.84(-) $&$<-6.46(-)$&$<-6.49(-)  $&	       &$<-7.04(-)  $&$<-6.83(-) $&	        &	     &	          & 	       &	    &$-6.96 $\\  
20& Ca      &$-5.68(05)$&$-5.55(08) $&$-5.53(14)  $&$-5.58(12)$&$-6.46(04)  $&$-6.36(03) $&$-5.74(07)  $&$-5.70(08) $&$-5.95(08) $&$-5.88(03) $&$-6.18(06) $&$-5.73 $\\  
21& Sc      &$-8.90(10)$&$-9.31(10) $&$-8.95(06)  $&$-8.88(11)$&$-9.39(01)  $&$-9.71(03) $&$-9.51(27)  $&$-9.44(08) $&$-10.06(01)$&$-10.90(-) $&$-9.79(01) $&$-8.99 $\\  
22& Ti      &$-7.19(-) $&$-7.18(-)  $&$-6.86(04)  $&$-6.89(07)$&$-6.95(04)  $&$-6.99(09) $&$-6.41(01)  $&$-6.72(05) $&$-7.08(04) $&$-6.81(30) $&$-7.01(07) $&$-7.14 $\\  
23& V       &           &            &$-7.48(05)  $&	       &$-7.60(-)   $&            &	        &$-7.19(05) $&	          &	       &$-7.16(-)  $&$-8.04 $\\  
24& Cr      &$-6.57(11)$&$-6.29(-)  $&$-6.01(07)  $&$-6.41(16)$&$-5.63(16)  $&$-5.65(08) $&$-5.93(10)  $&$-5.69(08) $&$-6.08(09) $&$-6.23(-)  $&$-5.79(13) $&$-6.40 $\\  
25& Mn      &           &            &$-6.32(01)  $&$-6.91(-) $&$-6.28(-)   $&$-6.31(03) $&$-6.28(-)   $&$-6.07(06) $&$-6.46(13) $&$-5.83(-)  $&$-6.26(01) $&$-6.65 $\\  
26& Fe      &$-4.72(12)$&$-4.75(10) $&$-4.30(07)  $&$-4.47(07)$&$-4.03(04)  $&$-4.14(05) $&$-4.13(10)  $&$-3.88(09) $&$-4.44(13) $&$-4.26(20) $&$-4.05(11) $&$-4.59 $\\  
27& Co      &           &            &             &	       &$-6.47(04)  $&$-6.15(-)  $&	        &$-6.12(12) $&	          &	       &$-6.16(04) $&$-7.12 $\\  
28& Ni      &$-5.71(15)$&$-5.56(01) $&$-5.53(03)  $&$-5.57(-) $&$-4.93(06)  $&$-5.14(07) $&$-5.10(15)  $&$-4.79(02) $&$-5.28(02) $&$-5.32(14) $&$-4.88(06) $&$-5.81 $\\  
29& Cu      &           &            &             &	       &$-6.93(-)   $&$-7.12(-)  $&	        &$-6.78(-)  $&	          &$-7.29(-)  $&	    &$-7.83 $\\  
30& Zn      &           &            &             &	       &$-6.65(-)   $&$-6.82(-)  $&$-7.29(-)   $&$-6.42(-)  $&	          &	       &$-6.59(-)  $&$-7.44 $\\  
39& Y       &$<-9.62(06)$&           &$<-9.78(-)  $&$<-9.02(16)$&$<-8.91(09)$&$<-8.98(05)$&$<-8.62(-)  $&$<-8.50(09)$&$<-8.77(13)$&$<-9.42(29)$&$<-8.76(08)$&$-9.83 $\\  
40& Zr      &           &            &             &	       &$-9.05(01)  $&$-8.51(01) $&	        &$-8.16(03) $&	          &	       &$-8.28(-)  $&$-9.45 $\\  
56& Ba      &$<-9.47(-)$&$<-9.95(-)  $&$<-9.13(04)$&$<-8.97(-) $&$<-8.50(-) $&$<-7.85(16)$&$<-8.09(18) $&$<-8.02(38)$&$<-8.97(09)$&$<-8.69(06)$&$<-7.77(10)$&$-9.87 $\\  
57& La      &           &            &             &           &$-9.61(09)  $&$-9.41(-)  $&	        &$-9.26(09) $&	          &	       &$-9.36(08) $&$-10.91$\\  
58& Ce      &           &            &             &           &$-9.26(04)  $&$-9.44(03) $&	        &$-8.94(15) $&	          &	       &$-9.14(-)  $&$-10.46$\\  
59& Pr      &           &            &             &           &             &            &	        &$-9.09(-)  $&	          &	       &	    &$-11.33$\\  
60& Nd      &           &            &             &           &$-9.48(03)  $&$-9.66(11) $&	        &$-9.23(07) $&	          &	       &$-9.41(04) $&$-10.59$\\  
62& Sm      &           &            &             &           &$-9.89(05)  $&            &	        &$-9.81(10) $&	          &	       &$-9.18(07) $&$-11.03$\\  
63& Eu      &           &            &             &           &$-10.20(-)  $&$-10.42(-) $&	        &            &	          &	       &$-9.95(-)  $&$-11.52$\\  
64& Gd      &           &            &             &           &             &   	  &	        &$-9.30(04) $&	          &	       &$-8.93(-)  $&$-10.92$\\  
68& Er      &           &            &             &           &$-9.65(-)   $&   	  &	        &$-9.39(-)  $&	          &	       &$-9.54(-)  $&$-11.11$\\  
70& Yb      &           &            &             &           &$-9.87(-)   $&   	  &	        &	     &	          &	       &	    &$-10.96$\\  
71& Lu      &           &            &             &           &             &            &             &$-10.20(-) $&	          &	       &$-10.3(-)  $&$-11.98$\\ 
\hline
&   \Teff   & 7660	& 7300       & 9382	   & 8450      & 7570	     & 8070       & 8170	& 8350       & 8020	  & 7232       & 8070	    & \\
&   \logg   & 4.02	& 2.92       & 3.78	   & 3.90      & 4.05	     & 3.97       & 4.00	& 4.15       & 3.69	  & 3.82       & 3.78	    & \\
&   \vmic   & 2.6	& 2.7	     & 1.9	   & 2.4       & 3.6	     & 2.6	  & 2.5	        & 2.9	     & 2.5	  & 2.8	       & 2.3	    & \\
&   \vmac   & 0.0	& 0.0	     & 0.0	   & 0.0       & 10.0	     & 0.0	  & 0.0	        & 0.0	     & 0.0        & 0.0	       & 0.0	    & \\
&   \vsini  & 73	& 127	     & 10	   & 73	       & 10	     & 29	  & 47	        & $<$5       & 62	  & 66	       & 10	    & \\
\hline
\end{tabular}
}
}
\end{table*}


In the following sections we comment on individual stars, devoting special 
attention to the elements that characterise the peculiarities of each star.
\subsection{Normal A-type stars}\label{normalstars}
\subsubsection{HD~73430}\label{hd73430}
\begin{figure}[ht]
\begin{center}
\resizebox{\hsize}{!}{\includegraphics{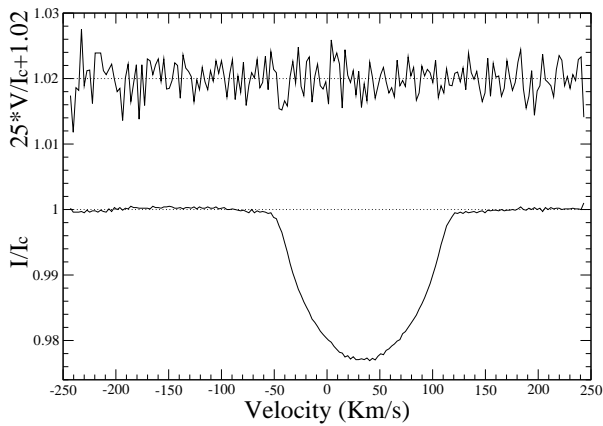}}
\caption{LSD profiles for the normalised Stokes $I$ and Stokes $V$ spectra
(from bottom to top) of HD~73430. The Stokes $V$ profile is expanded by a 
factor of 25 and shifted upward of 1.02.}
\label{LSD.HD73430}
\end{center}
\end{figure}

This is the slower rotator (\vsini\ = 73\,\kms) of the two normal A-type 
stars studied. The LSD profile (Fig.~\ref{LSD.HD73430}) shows symmetric 
rotational broadening, and does not show any particular features; the analysis 
gives a longitudinal magnetic field of $\langle B_z\rangle=1 \pm 45$ G. 

The effective temperature was determined so as to cancel any dependence of
abundance versus excitation potential for \ion{Fe}{i}. The \logg\, was 
calculated imposing equilibrium between the \ion{Fe}{i} and \ion{Fe}{ii} 
abundances. The values for \Teff\ and \logg\, found in this way 
were then averaged with the values found with the photometry. The 
\Teff\ was then checked and found consistent with the H$\alpha$ and H$\beta$ 
profiles, very sensitive to effective temperature at \Teff\ $<$ 8000\,K.

We find that C and O are solar and N is overabundant, as is observed for the 
other normal A-type star of our sample, HD~73575 (see Sect.~\ref{hd73575}). 
This differs from the Am stars, in which these elements are usually observed 
to be underabundant. We note that \citet{asplundetal2005} have significantly 
reduced the solar abundances of C, N and O, so all comparisons with the Sun of 
previous A-star studies regarding these elements may need to be revised. 
Ca, Sc, Ti, Cr, Fe and Ni are solar within the errors, allowing us to confirm 
the previous classification of HD~73430 as a normal A-type star.
\subsubsection{HD~73575}\label{hd73575}
\begin{figure}[ht]
\begin{center}
\resizebox{\hsize}{!}{\includegraphics{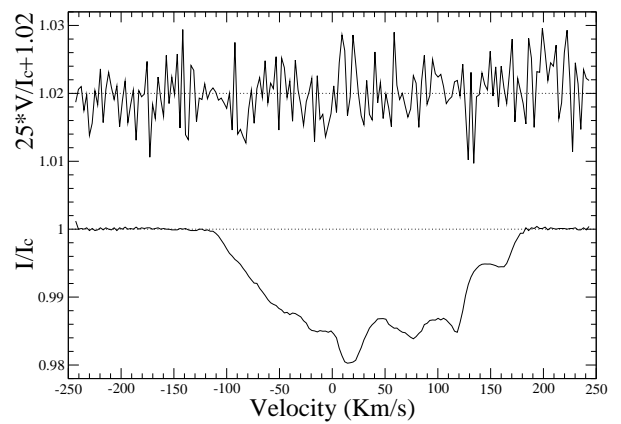}}
\caption{LSD profiles for the normalised Stokes $I$ and $V$ (from bottom to 
top) spectra of the $\delta$\,Scuti star HD~73575. The Stokes $V$ profile is 
expanded by a factor of 25 and shifted upward of 1.02.}
\label{LSD.HD73575}
\end{center}
\end{figure}

This is the faster rotator (\vsini\, = 127\,\kms) of the two normal A-type 
stars. HD~73575 is present in the catalog of \citet{rodriguezetal2000} of 
$\delta$\,Scuti variable stars. The Stokes $I$ LSD profile 
(Fig.~\ref{LSD.HD73575}) shows remarkable structure that may well be due to 
pulsation. No magnetic field was detected 
($\langle B_z\rangle=-215 \pm 149$\,G). 

The very high \vsini\, resulted in a relatively small number of metal lines 
with which to tune the fundamental parameters. The effective temperature was 
also determined by fitting the H$\alpha$ and H$\beta$ line profiles, which 
were then combined with the values obtained with the photometry. The \logg\, 
was determined by imposing equilibrium between different ionisation stages 
(\ion{Fe}{i} and \ion{Fe}{ii}). Due to the very high rotational velocity, the 
errors on the fundamental parameters are estimated to be similar to the 
photometric ones. 

HD~73575 shows all the properties of normal A-type stars of the cluster 
(e.g. C and O solar with N overabundant; Fe underabundant), except for 
Sc, which appears to be slightly underabundant. 
\subsection{The Blue Straggler HD~73666}\label{hd73666}
HD~73666 is the hottest star of the sample and the only Blue Straggler present
in the cluster, according to \citet{andri1998}. 
It is also included in the "New Catalogue of Blue Stragglers in Open Clusters"
by \citet{BS2007}.
The photospheric chemistry of this star has been previously investigated by 
several authors \citep{andri1998,burkcoupry1998}, but their results, the 
iron abundances for example, vary wildly ([Fe/H] = $+0.1$, $+0.5$ 
respectively). These discrepancies are probably due to the narrow wavelength 
ranges available to those investigators. 

\begin{figure}[ht]
\begin{center}
\resizebox{\hsize}{!}{\includegraphics{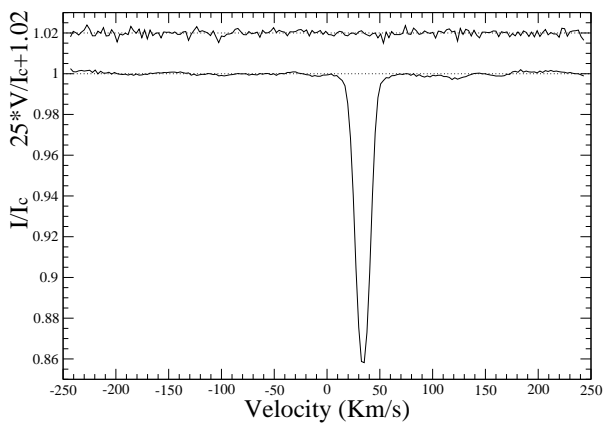}}
\caption{LSD profiles for the normalised Stokes $I$ and $V$ (from bottom to 
top) spectra for the Blue Straggler HD~73666. The Stokes $V$ profile is 
expanded by a factor of 25 and shifted upward of 1.02.}
\label{LSD.HD73666}
\end{center}
\end{figure}

The result of the application of the LSD technique is shown in 
Fig.~\ref{LSD.HD73666}.
The derived longitudinal magnetic field is $\langle B_z\rangle=6 \pm 5$\,G.

This star is the primary component of a close binary system (SB1), as is the
case for many other Blue Stragglers \citep{leonard1996}, 
and the flux coming from the secondary star is negligible, as previously 
checked by \citet{burkcoupry1998}, so the star was analysed ignoring the
presence of the secondary.

The sharp lines (\vsini\, = 10\,\kms) allowed for a detailed abundance 
analysis. The microturbulence velocity and \Teff\ were determined respectively 
by eliminating the abundance--equivalent width and abundance--excitation 
potential correlations, using \ion{Fe}{i}, \ion{Fe}{ii} and \ion{Ti}{ii} lines. 
The \logg\ value was obtained using the equilibrium between \ion{Fe}{i} and 
\ion{Fe}{ii}, and finally by looking for the best fit of the profile for the 
H$\alpha$ and H$\beta$ line wings (very sensitive to \logg\, at \Teff\, 
$>$ 8000K). The resulting \logg\ is the average of the values obtained with 
the two methods.

For the adopted model, \ion{Mg}{i} and \ion{Mg}{ii} deviate noticeably from 
the ionisation equilibrium condition. Their abundances are $-3.98$ and 
$-4.35$ respectively. The same behavior was found for \ion{Ca}{i} and 
\ion{Ca}{ii} with abundances of $-5.65$ and $-5.41$ respectively. These 
features of the Mg and Ca abundances explain the high value of the associated 
errors.
%

\begin{figure}[ht]
\begin{center}
\resizebox{\hsize}{!}{\includegraphics{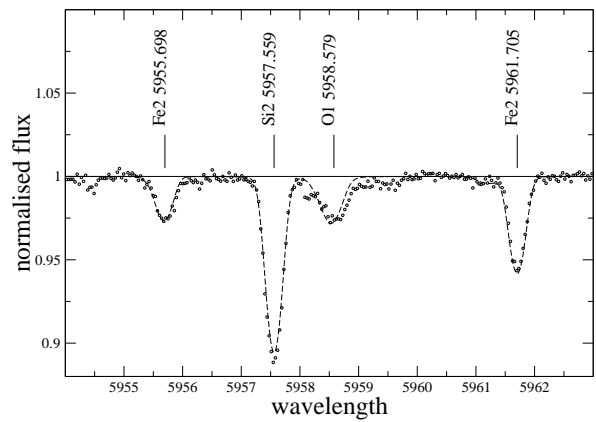}}
\caption{Portion of the spectrum of HD~73666. The dots show the observation, 
the dashed line shows the synthetic spectrum. The weak blended lines are 
omitted.}
\label{si}
\end{center}
\end{figure}

\citet{abt1985} classified this star as Ap\,(Si). The signature of this kind 
of star is a strong silicon overabundance. However, it can be seen in 
Table~\ref{abbondanze del sample} and Fig.~\ref{abbondanze per hd73666} that 
the Si abundance is not much higher than the solar abundance of this element. 
Also Ap\,(Si) stars are magnetic, and the very high precision of the 
longitudinal field diagnosis effectively rules out the presence of any 
organised magnetic field. Fig.~\ref{si} shows one of the selected Si lines. 
If HD~73666 was an Ap(Si) star, we would expect this line to be much stronger, 
compared to the surrounding Fe lines.
\citet{andri1998} cataloged this star as an Am star, but Ca is slightly 
overabundant while Sc is solar. 
C, N and O are overabundant and, as it is possible to see in 
Fig.~\ref{abbondanze di tutti}, these overabundances are typical of the normal 
A-type stars of the Praesepe cluster. These results allow us to conclude that
HD~73666 is neither an Ap nor an Am star.
%
\subsection{Am stars}\label{Amstars}
In this section we analyse the stars of the sample classified as Am stars in 
previous works. According to \citet{preston} a star in the 
temperature range of 7000 -- 10000 K is classified as Am (CP1) if 
Ca and/or Sc are underabundant and the heavy metals are overabundant. 
Am stars nearly always show low values of \vsini, tipically less than 100 \kms.
\subsubsection{HD~72942}\label{hd72942}
\begin{figure}[ht]
\begin{center}
\resizebox{\hsize}{!}{\includegraphics{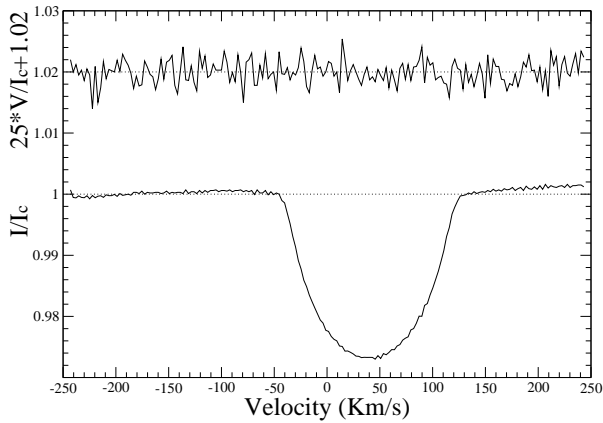}}
\caption{LSD profiles of the normalised Stokes $I$ and $V$ (from bottom to 
top) spectra of HD~72942. 
The Stokes $V$ profile is expanded by a factor of 25 and shifted upward of 
1.02.}
\label{LSD.HD72942}
\end{center}
\end{figure}

The LSD profile for HD~72942 is symmetric and rotationally broadened, as is 
illustrated in Fig.~\ref{LSD.HD72942}.
The magnetic analysis of the LSD profiles gives a value of the longitudinal 
magnetic field of $\langle B_z\rangle=12 \pm 31$\,G.

\Teff\ was set using \ion{Ti}{ii} and \ion{Fe}{i} lines. The \logg\ value was 
estimated using the equilibrium between the \ion{Fe}{i} and \ion{Fe}{ii} 
abundances and fitting the H$\alpha$ and H$\beta$ line profiles. 

The large error associated with the Y abundance is due to the poor 
determination of the line parameters and to the unknown \nlte\, effects 
associated with each line. It was not possible to derive the K abundance from 
the \ion{K}{i} line at 7698.974\,\AA\, because the line was blended by a 
telluric line. 

Ca and Sc are not underabundant, indicating that HD~72942 is probably not an 
Am star. However, the Fe abundance is similar to that observed in the metallic 
stars of the sample. 

HD~72942 is not known to be a spectroscopic binary and is neither a classical 
Am star nor a normal A-type star (Sect.~\ref{sezione,abn,tutti}). The 
calculated radial velocity is rather high as compared to those of the other 
members and to the mean of the cluster. 
See also the discussion in Sect.~\ref{sezione,abn,tutti}.
\subsubsection{HD~73045}\label{hd73045}
HD~73045 is the primary component of an SB1 binary system 
\citep{debernardietal2000}. It exhibits a low projected rotational velocity 
(\vsini\, = 10\,\kms). 

\begin{figure}[ht]
\begin{center}
\resizebox{\hsize}{!}{\includegraphics{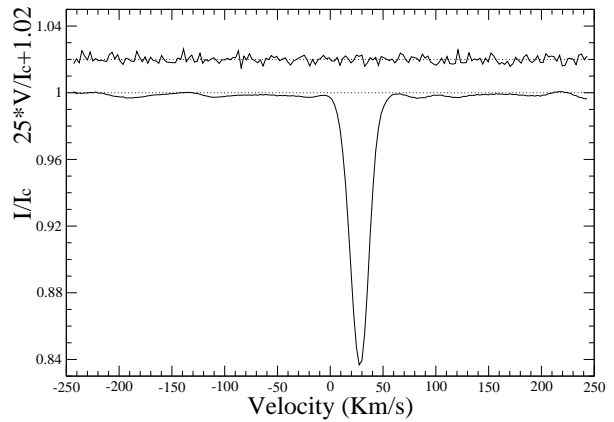}}
\caption{LSD profiles for the normalised Stokes $I$ and $V$ (from bottom to 
top) spectra of HD~73045. 
The Stokes $V$ profile is expanded by a factor of 25 and shifted upward of 
1.02.}
\label{LSD.HD73045}
\end{center}
\end{figure}

The LSD profiles of the Stokes $I$ and $V$ spectra are shown in 
Fig.~\ref{LSD.HD73045}. 
The magnetic field analysis provided a longitudinal magnetic field of 
$\langle B_z\rangle=-1 \pm 4$\,G.

The secondary star does not contribute significantly to the lines of the 
spectrum, according to \citet{debernardietal2000} and \citet{budaj1996}, so we 
have analysed the system as a single star. During our abundance determination 
we noticed an extra broadening of the wings of deep lines, perhaps due to the 
binarity. To be able to fit these lines we introduced a macroturbulence 
velocity (\vmac\, = 10 \kms\, - obtained fitting several deep lines) to 
compensate for this effect. The use of the macroturbulence velocity does not
lead to any significant abundance change, as expected. In particular the Iron 
abundance calculated without the use of the macroturbulence velocity results 
to be 0.01 dex lower than the final adopted one. The microturbulence velocity 
and the effective temperature were established using \ion{Fe}{i}, \ion{Fe}{ii} 
and \ion{Ni}{i} unblended lines, using the method described previously. The 
\logg\ value was obtained imposing the ionisation equilibrium condition 
for Fe. The determined \Teff\, was then checked with the H$\alpha$ and 
H$\beta$ line wings.

HD~73045 is a typical Am star as Ca and Sc are underabundant and the Fe-peak 
elements are overabundant, as are the rare earth elements. 
C, N and O are underabundant, typical of all the Am stars of the sample. 
For the \ion{Cu}{i} line at 5105.537\,\AA, \ion{Co}{i} line at 5342.695\,\AA\ 
and \ion{Co}{i} line at 5352.045\,\AA\ we applied a hyperfine structure 
correction, but the \vsini\ was too high to produce any detectable difference 
in the line broadening.
\subsubsection{HD~73730}\label{hd73730}
HD~73730 is an Am star with an intermediate rotational velocity 
(\vsini\ = 29\,\kms).

\begin{figure}[ht]
\begin{center}
\resizebox{\hsize}{!}{\includegraphics{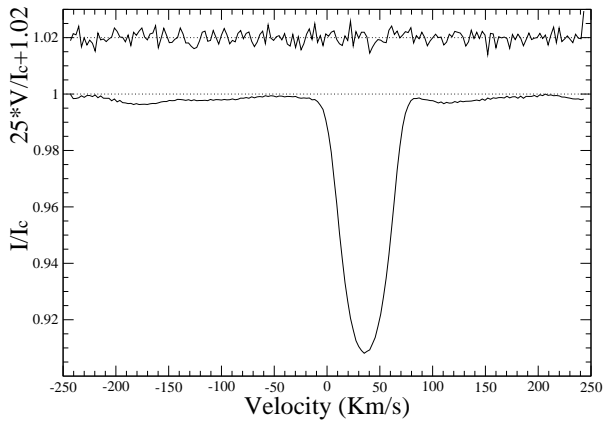}}
\caption{LSD profiles of the normalised Stokes $I$ and $V$ (from bottom to 
top) spectra of HD~73730. 
The Stokes $V$ profile is expanded by a factor of 25 and shifted upward of 
1.02.}
\label{LSD.HD73730}
\end{center}
\end{figure}

The LSD profiles for the Stokes $I$ and Stokes $V$ spectra are shown in 
Fig.~\ref{LSD.HD73730}.
No magnetic field was detected from the LSD profiles. The calculated 
longitudinal magnetic field was $\langle B_z\rangle=-12 \pm 9$\,G.

We obtained the fundamental parameters (\vmic, \Teff, and \logg) 
using only \ion{Fe}{i} and \ion{Fe}{ii} lines, since the \vsini\ did not 
allow us to select a sufficient number of unblended lines with good line 
parameters for any other element.
Since we had the possibility to use only \ion{Fe}{i} and \ion{Fe}{ii} lines, we 
also used the values of the \logg\ obtained from the fitting of the H$\alpha$ 
and H$\beta$ line profiles.

The \vsini\ and the non-binarity allowed us to perform a precise 
abundance analysis for many elements - in fact only C shows a large error bar.
This is probably due to the small number of selected C lines, all of which
present \nlte\ effects.
The underabundances of Ca and Sc confirm that HD~73730 is an Am star. 
Also C, N and O are underabundant, as is the case for the other Am stars of the
sample.

The \vsini\ is sufficiently large to hide any hyperfine structure broadening.
\subsubsection{HD~73618}\label{hd73618}
HD~73618 is the primary component of an SB1 binary system in which the flux of
the secondary star does not significantly influence the total flux spectrum, 
according to \citet{masonetal1993}.
For this star, as for all the others observed with the ELODIE spectrograph, we
do not have a Stokes $V$ LSD profile since the instrument did not support a 
polarimetric mode. So we checked for the presence of a magnetic field with the
correlation between abundance and Land\'{e} factor for \ion{Fe}{i}. This method 
allows the detection of magnetic fields stronger than about 1--2 kG 
\citep{ch2006}. We did not find any detectable magnetic field for HD~73618.
To obtain the \Teff\ and the \logg\ we used \ion{Fe}{i} and \ion{Fe}{ii} 
lines. The fitting of the H$\alpha$ and H$\beta$ lines was not applied 
because their wings were broader than the wavelength range of the respective 
orders of the spectrograph.
Since the \vsini\, is too high to get equivalent widths for enough 
unblended \ion{Fe}{i} lines, we used the \vmic\, calculated with 
Eq.~(\ref{vmicphotometry}).

Only Sc is underabundant, while Ca is solar. C and O are underabundant,
as we have found for the other Am stars of the sample. Also, the Fe-peak 
elements show a behavior typical of this type of star. This confirms the 
previous classification of HD~73618.
Because of the lower resolution of the ELODIE spectrograph with respect to the 
\espa\ spectrograph, the errors associated with the abundances turn out to be 
higher than those calculated for the other stars described above.

This star is included, as a Blue Straggler, in the catalogue by \citet{BS2007}.
\citet{andri1998} analysed this star in his
paper on Blue Stragglers in the Praesepe cluster, but he did not define it 
as a true Blue Straggler, since in the colour--magnitude diagram it does not 
lie outside the cluster's main sequence. Our sample of stars is too small to 
allow us to generate a trustworthy colour--magnitude diagram based on 
spectroscopic temperatures. For this reason we are not able to confirm either 
the classification of \citet{BS2007} or that of \citet{andri1998}.
\subsubsection{HD~73174}\label{hd73174}
HD~73174 is the primary component of a triple SB1 system
\citep{debernardietal2000} and is the slowest rotator of the program stars, 
with \vsini\, $<$ 5\,\kms. The spectrum was obtained with the ELODIE 
spectrograph with a mean resolution of 45\,000, not sufficient to allow an 
accurate measurement of the \vsini\ of this star.

\begin{figure}[ht]
\begin{center}
\resizebox{\hsize}{!}{\includegraphics{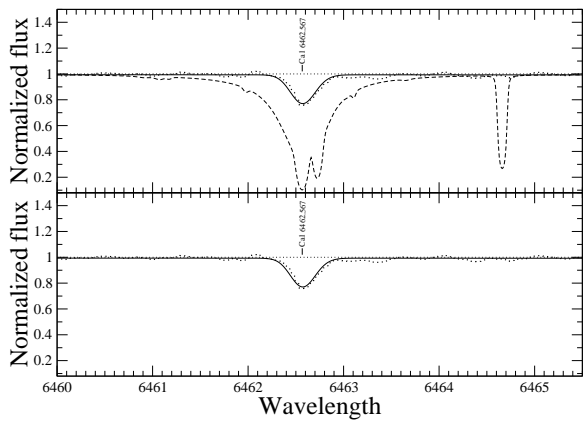}}
\caption{Comparison between the incoming flux from the two components of the
HD~73174 system. In the upper panel the dotted thick line is the 
observed (normalised) spectrum, while the solid and the dashed lines 
are the synthetic spectra of the primary and the secondary star respectively. 
These two synthetic spectra are calculated with the parameters described in 
the text and with solar abundances. In the lower panel we show with the dotted 
thick line the observed and normalised spectrum and with the straight line 
the resultant synthetic spectrum after the application of the calculated radius 
ratio ($R_{1}/R_{2}$ = 3.26).} 
\label{dimostrazione secondaria piccola} 
\end{center} 
\end{figure}

\cite{debernardietal2000} obtained the masses of the three components of the 
system (2 M$_{\sun}$; 0.68 M$_{\sun}$; 0.63 M$_{\sun}$). Knowing the age of 
the cluster \citep[$\log(t) = 8.85$]{gonzalez} it was possible to 
estimate the effective temperatures and the radii of the three stars 
comprising the system. Using the evolutionary mass models of 
\citet{schaller1992}, we estimated the effective temperature and the radius of 
the primary component (8375 K and 2.22 R$_{\sun}$ respectively). From plots on 
evolutionary models of cool stars of \citet{gray1992} we estimated the 
effective temperature and the radii of the other two components: 4557 K and 
0.68 R$_{\sun}$ for the secondary and 4300 K and 0.65 R$_{\sun}$ for the third 
component. Since the presence of the third component is not certain we have 
not taken this star into account in our analysis. The radius ratio between the 
primary and the secondary star turns out to be $R_{1}/R_{2}$ = 3.26. We 
assumed a \vmic\, of 2\,\kms\ for both the stars and a \logg\, of 4.2 and of 
4.4 for the primary and the secondary respectively. We calculated a model 
atmosphere for each of the two stars, using the parameters just described, 
and produced a synthetic spectrum for each of the two stars to check how much 
flux in the observed intensity spectrum is due to the secondary component. As 
it is possible to see in Fig.~\ref{dimostrazione secondaria piccola}, the flux 
coming from the secondary star is negligible with respect to that of 
HD~73174. For this reason we analysed the system as a single one.
The error bars on these inferred fundamental parameters could be very high 
since the error bars on the mass determination are unknown.

The low \vsini\ allowed us to perform a very detailed abundance analysis and 
the errors associated with the various elements are for this reason quite low, 
even if the signal to noise ratio and the resolution were not optimal.

C, N, O and Sc are underabundant while Ca is almost solar, as was found for 
HD~73618. For this reason we confirm the previous classification of this star.
Only Ba shows a large error bar. This is probably due to the strong 
difference in depth of the selected lines and to the different influence that
the \nlte\, effects have on each of these lines.

We took into account the hyperfine structure correction for the following 
lines: \ion{Co}{i} at 5342.695\,\AA\, and 5352.045\,\AA, \ion{Cu}{i} at 
5105.537\,\AA, \ion{Mn}{i} at 6021.819\,\AA\, and \ion{Zn}{i} at 
4722.153\,\AA. The abundance correction from HFS observed for all of these 
lines is smaller than 0.03 dex, much less than our typical errors.

We have to comment the difference of about 700 K between the \Teff\, 
derived from Str\"{o}mgren photometry and from spectroscopy. We are not able 
to explain the reason for this dicrepancy. Probably, give concerning the very 
low \vsini\, of this star, a spectrum taken with a much higher resolution 
could solve this puzzle. 
\subsubsection{HD~73711}\label{hd73711}
The ELODIE spectrum of this Am star has unfortunately a low SNR 
(Table~\ref{tabella radec}) and a quite high \vsini\ (62\,\kms), making the 
abundance analysis more difficult and less precise. 
HD~73711 is also the primary component of an SB1 system. The abundances of 
this star were previously analysed by \citet{burkcoupry1998} who considered 
the spectrum as that of a single star. 

We have used \ion{Fe}{i} and \ion{Fe}{ii} lines to calculate the \Teff\ and 
\logg, while \vmic\ was calculated with Eq.~(\ref{vmicphotometry}). 
As with all the other ELODIE spectra we were not able to use the H$\alpha$ 
and H$\beta$ lines to determine the parameters. We have not found any 
detectable magnetic field with the abundance-Land\'{e} factor correlation
for \ion{Fe}{i} lines.

We confirm the previous classification of HD~73711, as Ca and Sc are
underabundant, like C and O, as occurs for all the Am stars of the sample.
The errors associated with the abundances are below 0.15 dex in spite of the 
high \vsini\ and the low SNR; only the error associated with the C abundance 
reaches 0.15 dex, probably due to the small number of selected lines used to 
calculate the final abundance, which were analysed without taking into account 
\nlte\ effects. 
\subsubsection{HD~73818}\label{hd73818}
The spectrum of the Am star HD~73818 has the lowest SNR of the whole sample. 
This fact, combined with the rather high \vsini\, (66\,\kms), made the 
abundance analysis very difficult and imprecise. This is the primary star of 
an SB1 system. We analysed this spectrum as a single star, as 
suggested by \citet{burkcoupry1998}.

We tried to calculate \Teff\ and \logg\ using \ion{Fe}{i} and
\ion{Fe}{ii} lines, applying the method described in Sect.~\ref{howparameters}, 
but we were never able to converge to two final values. For this reason we 
adopted the fundamental parameters calculated with the photometry. The \vmic\ 
was calculated using Eq.~(\ref{vmicphotometry}).

The abundances of many elements show large errors, mainly those elements for 
which \nlte\ effects should be taken into account. The error associated with 
the Fe abundance is quite high in comparison to all other stars 
of the sample. This example illustrates the importance of a high SNR 
to be able to perform an accurate abundance analysis that 
starts with a spectroscopic determination of the fundamental parameters.

Ca and Sc are underabundant, confirming in this way the previous 
classification. Only the C abundance is higher than expected for 
an Am star of the Praesepe cluster. 
%
\subsubsection{HD~73709}\label{hd73709}
HD~73709 is the primary star of an SB1 quadruple system \citep{abt1999b}.
The low \vsini\, allowed a good determination of the elemental
abundances; only the moderate resolution of MUSICOS decreased the quality of 
our results.

This star was classified as Am by \citet{gray1989}, but was found
photometrically to be an Ap star by \citet{maitzen1987} according to the
$\Delta a$ index ($\Delta a$ = $+0.018$ mag). \citet{debernardietal2000},
using two ELODIE spectra and observations with the
CORAVEL radial velocity scanner, found a magnetic field of 7.5 kG. 
\citet{shorlin2002} looked for the presence of such a
magnetic field using the LSD technique applied to the same \mus\ spectrum 
we analyse here. Their result shows clearly that HD~73709 
does not have a significant magnetic field, with the measured value equal to 
$\langle B_z\rangle=66 \pm 39$\,G.

We calculated a synthetic $\Delta a$ value with respect to the theoretical 
normality line $a_0$ determined by \citet[][Sect.~3.3.2]{sergey2007} and found 
that $\Delta a = +0.012$\,mag, which is compatible with the observed value.

As mentioned before, this star is the main component of a quadruple system. We
ruled out that the third or fourth components could contribute significantly to
the light since they appear to be much less massive than the other components. 
The ratio between the masses of the primary and secondary component is 
$M_{2}/M_{1}$ $\geq$ 0.27 \citep{abt1999b}. For HD~73174 we employed a value 
of $M_{2}/M_{1}$ $\geq$ 0.35 \citep{abt1999b} and in that case our test 
demonstrated that the flux coming from the secondary star was negligible with 
respect to the primary for a radius ratio lower than two (Sect.~\ref{hd73174}).
For this reason we arrived immediately at the same conclusion as for HD~73174 
- that the contribution of the secondary to the observed spectrum is negligible.

The \vmic\ and \Teff\ were established using \ion{Fe}{i}, \ion{Fe}{ii} 
and \ion{Ni}{i} unblended lines. The \logg\ was set imposing the ionisation 
equilibrium condition for \ion{Fe}{i} and \ion{Fe}{ii}. The fitting of the 
H$\alpha$ and H$\beta$ lines was not applied to obtained the fundamental 
parameters, for the same reason as for the ELODIE spectrograph.

O, Ca and Sc are underabundant, confirming in this way the previous 
classification. C and Al show large error bars ( $\geq$ 0.15\,dex), probably 
due to the fact that \nlte\ effects were not taken into account.  
\section{Discussion}\label{discussion}
\subsection{Are weak magnetic fields present in the Am stars of Praesepe?}\label{am as magnetic}
If magnetic fields are present in Am stars, this might help explain some of 
the peculiar properties of these stars, such as slow rotation and the typical 
Ca/Sc underabundances \citep{bohm}. 

We decided to use the LSD approach to 
detect magnetic fields in Am stars since this method is the most precise 
method currently available, especially for stars with rich line spectra and 
low \vsini. As explained in Sect.~\ref{howparameters} we have also looked 
for magnetic fields using other techniques, but these methods have resulted 
in substantially higher upper limits.

\citet{lanzmathys} have claimed evidence of magnetic fields in the Am star 
$o$~Peg (\Teff\, = 9550) using the difference of equivalent widths of two 
\ion{Fe}{ii} lines at 6147.741 \AA\ and 6149.258 \AA. We did not try to apply 
the same method to compare our results to theirs as all of the sharp-lined Am 
stars in our sample are spectroscopic binaries, and the blending due to the 
spectral lines of the secondary might well affect the subtle signal of a 
magnetic field. Another problem is due to the fact that our Am stars are cool 
Am stars and the \ion{Fe}{ii} line at 6147.741 \AA\ is strongly blended by an 
\ion{Fe}{i} line, preventing an accurate equivalent width determination.

\citet{shorlin2002} analysed 25 Am stars with the LSD technique without 
finding any significant magnetic field signature. These authors concluded that 
no credible evidence exists for the presence of organised or complex magnetic 
fields in Am or normal A-type stars.

In our study, the three Am stars observed with \espa\ and the one observed 
with \mus\ did not reveal any magnetic field signature in the circular 
polarized spectra, analysed with the LSD procedure. We also tried to detect 
the presence of potential magnetic fields in the other Am stars, observed with 
ELODIE, using the method described in Sect.~\ref{howparameters}. For all of 
them it was not possible to detect any magnetic field. 

Our conclusion is that the cool Am stars of Praesepe (\Teff\,$<$ 8500 K) show 
no evidence of magnetic fields that could explain the well-known phenomena 
associated with Am stars. This result supports and strengthens the conclusion 
of \citet{shorlin2002}.
\subsection{Normal vs. Am stars}\label{sezione,abn,tutti}
\begin{figure*}[ht]
\begin{center}
\rotatebox{270}{
\includegraphics[width=230mm]{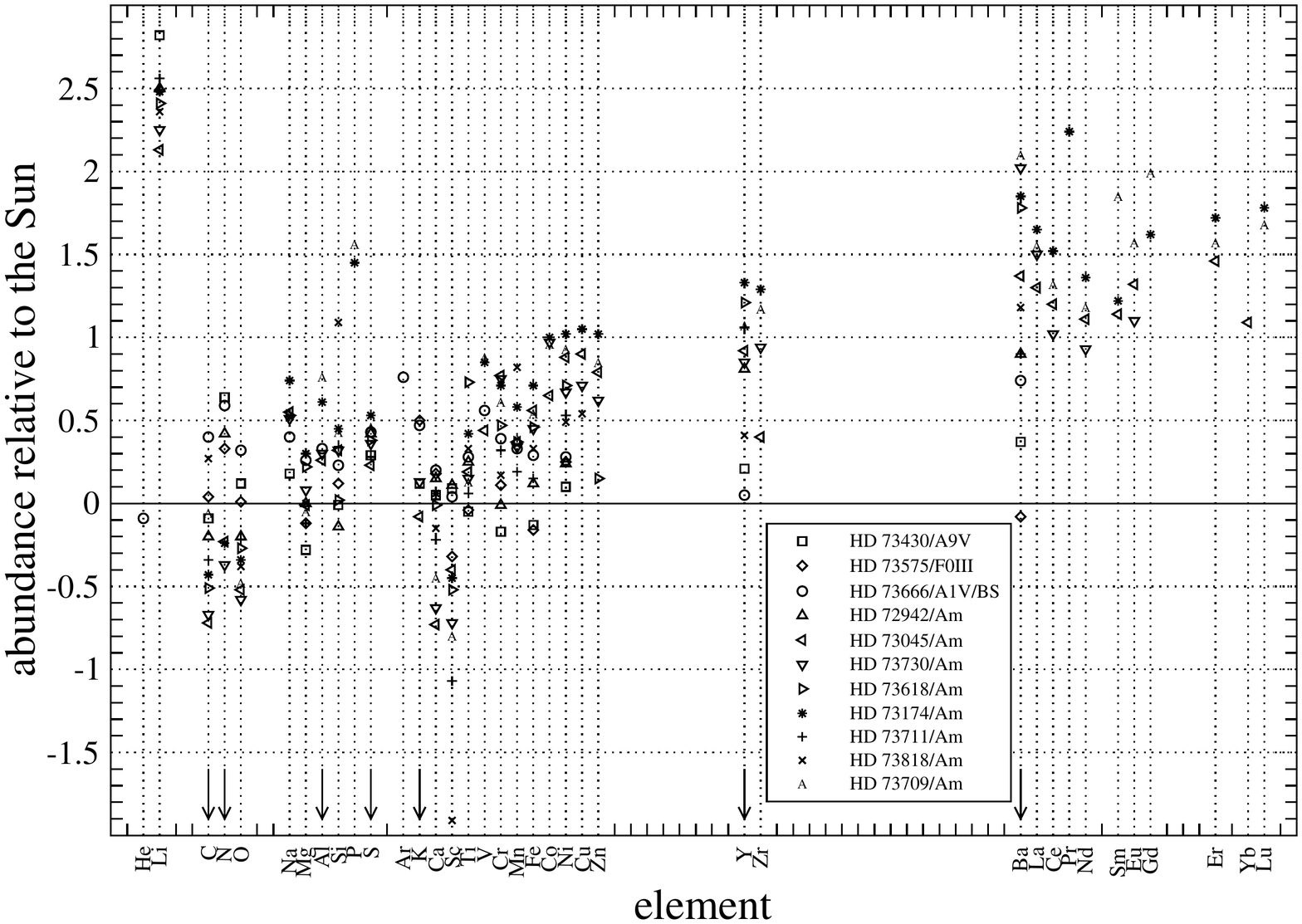}
}
\caption{Elemental abundances relative to the Sun for the program stars. 
The solar abundances are taken from \citet{asplundetal2005}. In order to 
show a more readable Figure, the errors (Table~\ref{abbondanze del sample}) 
are omitted. Arrows show elements affected by \nlte\, effects.}
\label{abbondanze di tutti}
\end{center}
\end{figure*}

In Fig.~\ref{abbondanze di tutti} and in Table~\ref{abbondanze del sample} 
we show the final adopted abundances for all of the program stars.  
The precision obtained with the abundance analysis allows us to distinguish 
two groups among the A-type stars of Praesepe: the normal A-type stars and the 
Am stars. For the two kinds of star analysed in this paper, the abundances of 
the Fe-peak elements are different. In normal A-type stars, Fe, Ni, and Cr 
are solar or slightly underabundant (compared to their solar abundances). In 
Am stars, these elements are slightly overabundant. In all but one of the 
stars of our sample, Ba is overabundant. However, the Ba abundance is 
definitely lower in normal A-type stars than in Am stars. This suggests that 
Ba may be considered as an additional indicator of Am peculiarities, together 
with C, N, O, Ca, Sc and the Fe-peak elements.

The abundance pattern found for HD~72942 deserves further comment. 
\citet{bidelman} classified HD~72942 as Am. Our abundance analysis shows some 
features common to the Am stars, like underabundances of C and O and an 
overabundance of Fe, and some others in common with normal A-type stars, like 
overabundance of N, Ca and Sc and a low Ba abundance. For these reasons we 
believe that HD~72942 has a chemical composition between the normal A-type 
stars and the Am stars. The high observed radial velocity of this star may be 
indicative of undetected binarity.

The remaining Am stars represent a fairly homogeneous sample. 

\subsection{Constraints to the diffusion theory}
Underabundances of Sc and/or Ca and small overabundances of Iron-group 
elements are qualitatively explained by diffusion theory (see, e.g., the 
review in the introduction by \citet{alecian} and references therein). In 
particular, following \citet{alecian}, \citet{richer2000} developed a detailed 
modeling of the structure and evolution of Am/Fm stars, taking into account 
atomic diffusion of metals and radiative accelerations for all species in the 
OPAL opacities.

\begin{figure}[ht]
\begin{center}
\includegraphics[width=100mm]{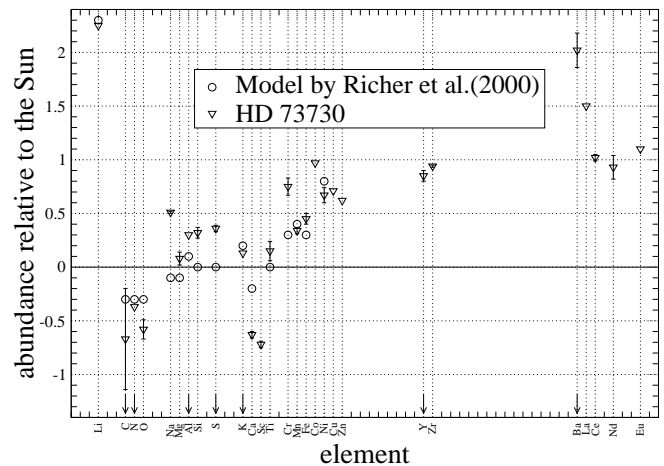}
\caption{The open triangles show the abundances derived for HD~73730. The open
circles show the predicted abundances displayed in Fig.~14 of
\citet{richer2000}. Arrows show elements affected by \nlte\, effects}
\label{abbondanze per hd73730}
\end{center}
\end{figure}

The results of this work can be used to constrain diffusion models, and we can 
compare our results with theoretical predictions. For the comparison, we use 
the abundances of HD~73730 (Fig. \ref{abbondanze per hd73730}) since it is the 
star among those that are not binary with the lowest \vsini, which favors the 
abundance analysis. Furthermore, the elemental abundances of this star are also
representative of the homogeneous group of Am stars of the Praesepe cluster as 
listed in Sect.~\ref{sezione,abn,tutti}.

Figure 14 of \citet{richer2000} shows the predicted abundances from a
diffusion and turbulence model as a function of time and effective
temperature. Open circles (reported also in Fig~\ref{abbondanze per hd73730}) 
refer to the predicted abundances for 670 Myr old stars, an age that is 
closely comparable to that one of the Praesepe cluster 
\citep[$\sim$700 My, according to][]{gonzalez}. For temperatures around 
\Teff\,$= 8000$\,K, the temperature of HD~73730, the 
model by \citet{richer2000} predicts underabundances of -0.3 (relative to the 
Sun) for C, N and O, of $-0.1$ for Na, Mg and Ca, solar abundances for Si, 
S and Ti, overabundances of $+0.1$ for Al, of $+0.3$ for Cr and Fe, of $+0.4$ 
for Mn, of $+0.8$ for Ni and $+2.3$ for Li, within 0.1 dex. 

Our Fig.~\ref{abbondanze di tutti} (see also 
Table \ref{abbondanze del sample}) shows an agreement for Li, C, N, O, Mg, 
Al, Si, Ca and Fe-peak elements within 0.1 dex.
Discrepancies are apparent for Na and S (Sc was not modeled by 
\citet{richer2000} because no OPAL data was available for Sc).


Na is the element for which the derived abundance ($\simeq 0.5$\,dex with 
respect to the solar abundance) has the smallest scatter among the Am stars 
of our sample. Observations seems definitely to suggest an overabundance of 
this element which is inconsistent with model predictions of $-0.1$ dex.

S shows a small scatter in our sample ($\sigma$ = 0.12 dex).
\citet{kamp2001} analysed the S abundance in the metal poor A-type star
$\lambda$ Boo in LTE and \nlte. Their analysis shows that the S abundance
determined in \nlte\, is lower than that determined in LTE by 0.1
dex. For metal poor stars \nlte\, effects are larger than for normal A-type
stars and Am stars. This brings our S abundance ($\simeq +0.35$ dex) to an 
overabundance (relative to the Sun) of $\geq$0.25 dex, as compared with a 
solar abundance predicted by \citet{richer2000}. This apparent discrepancy 
should be investigated in \nlte\, regime. 


%
\subsection{Lithium abundance}\label{li}
In the context of radiative diffusion theory, it is interesting to examine the
atmospheric abundance of lithium in stars where such a mechanism is known to be
present, like in Am stars. Such studies have been carried out especially by
\citet{burkcoupry1991} and \citet{burkhart2005}.
Their conclusion was that, in general, the Li abundance in
Am stars is close to the cosmic value ($\log{N_{Li}/N_{total}} \sim 
-9.04$~dex), although a small proportion of them are Li deficient. The normal
A-type stars appear to have a higher Li abundance 
($\log{N_{Li}/N_{total}} \sim -8.64$~dex), in the
\Teff\, range 7000--8500 K \citep{burkhart1995}.

\begin{figure}[ht]
\begin{center}
\includegraphics[width=95mm]{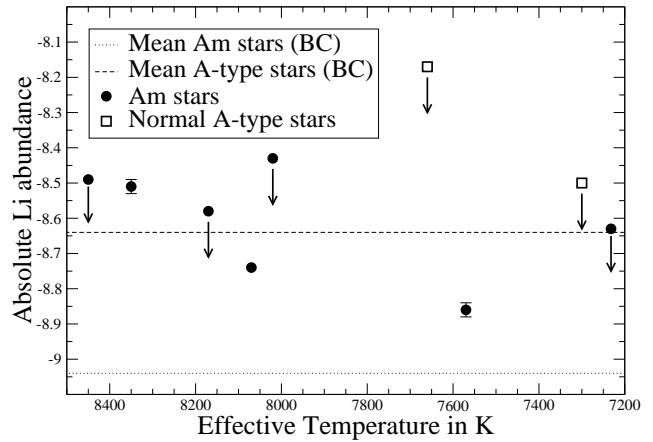}
\caption{Distribution of the Li abundance as a function of the 
\Teff. The dashed line shows the Li abundance value found by 
\citet{burkhart1995} (BC in the legend) for normal A-type stars. The 
dotted line shows their Li abundance value for Am stars. The open 
squares and the closed circles correspond to the Li abundances of 
the program normal A-type stars and Am stars respectively. The upper limit
values are indicated with an arrow.} 
\label{livsteff} 
\end{center} 
\end{figure}

\begin{figure}[ht]
\begin{center}
\resizebox{\hsize}{!}{\includegraphics{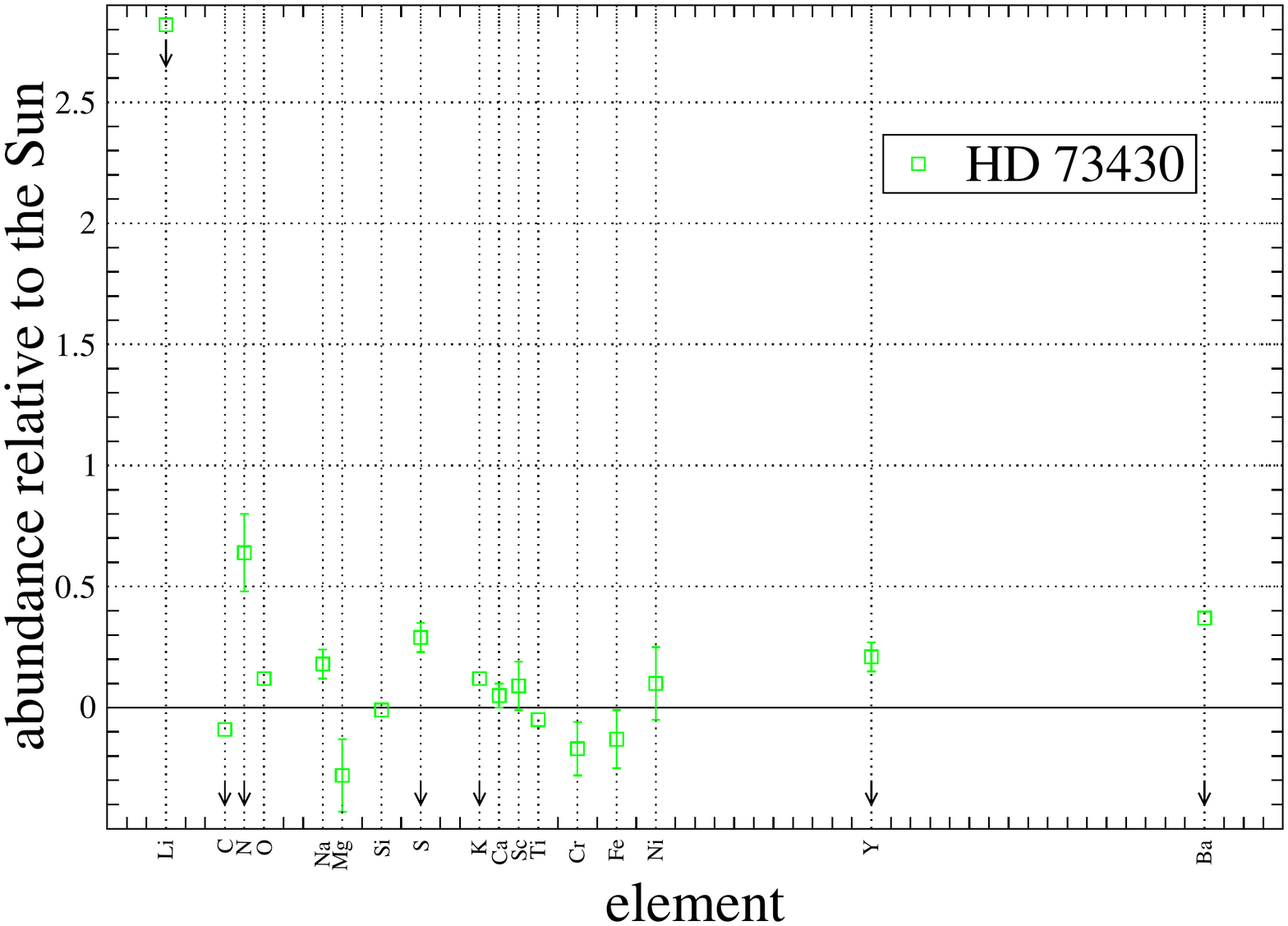}}
\caption{Elemental abundances relative to the Sun for HD~73430. The solar
abundances are taken from \cite{asplundetal2005}. Arrows show elements
affected by \nlte\, effects. Errors are given in 
Table\ref{abbondanze del sample}.}
\label{abbondanze per hd73430}
\end{center}
\end{figure}

\begin{figure}[ht]
\begin{center}
\resizebox{\hsize}{!}{\includegraphics{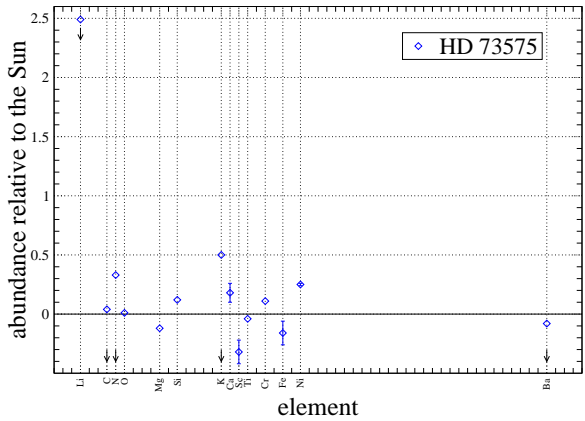}}
\caption{As for Fig.~\ref{abbondanze per hd73430} for HD~73575}
\label{abbondanze per hd73575}
\end{center}
\end{figure}

\begin{figure}[ht]
\begin{center}
\resizebox{\hsize}{!}{\includegraphics{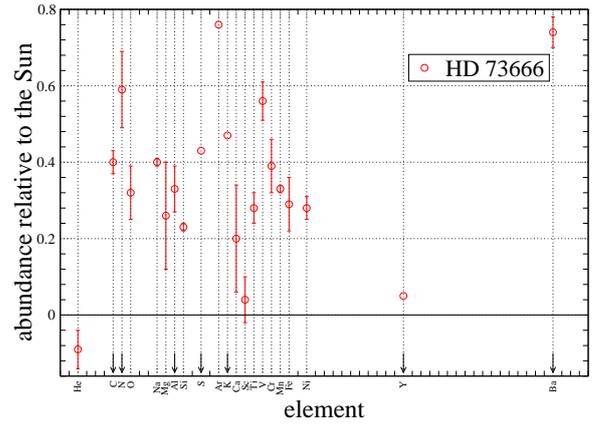}}
\caption{As for Fig.~\ref{abbondanze per hd73430} for HD~73666}
\label{abbondanze per hd73666}
\end{center}
\end{figure}

\begin{figure}[ht]
\begin{center}
\resizebox{\hsize}{!}{\includegraphics{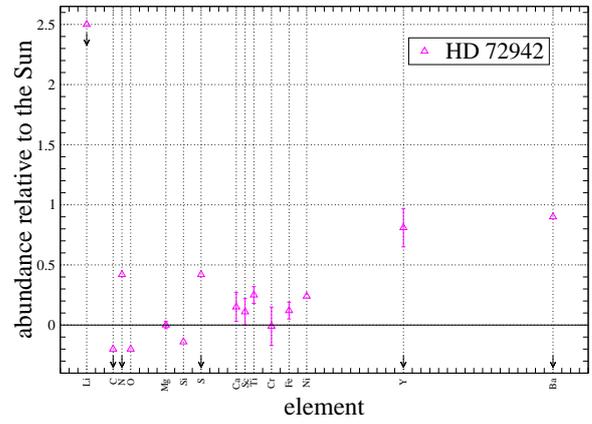}}
\caption{As for Fig.~\ref{abbondanze per hd73430} for HD~72942}
\label{abbondanze per hd72942}
\end{center}
\end{figure}

\begin{figure}[ht]
\begin{center}
\resizebox{\hsize}{!}{\includegraphics{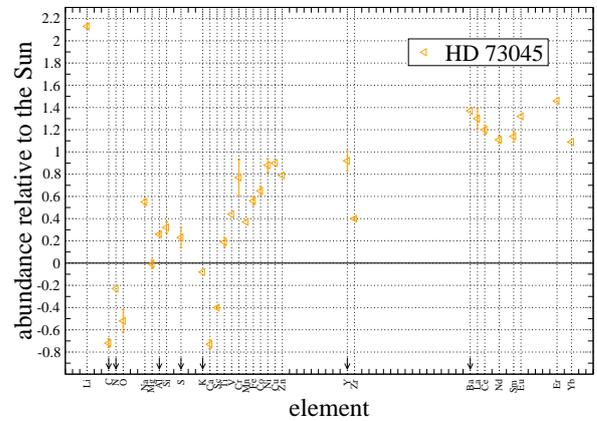}}
\caption{As for Fig.~\ref{abbondanze per hd73430} for HD~73045}
\label{abbondanze per hd73045}
\end{center}
\end{figure}

\begin{figure}[ht]
\begin{center}
\resizebox{\hsize}{!}{\includegraphics{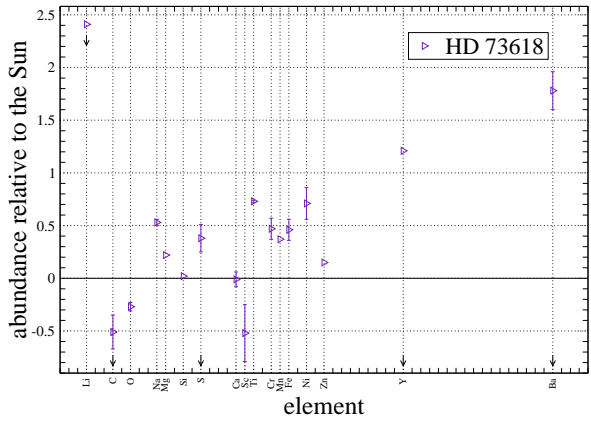}}
\caption{As for Fig.~\ref{abbondanze per hd73430} for HD~73618}
\label{abbondanze per hd73618}
\end{center}
\end{figure}

\begin{figure}[ht]
\begin{center}
\resizebox{\hsize}{!}{\includegraphics{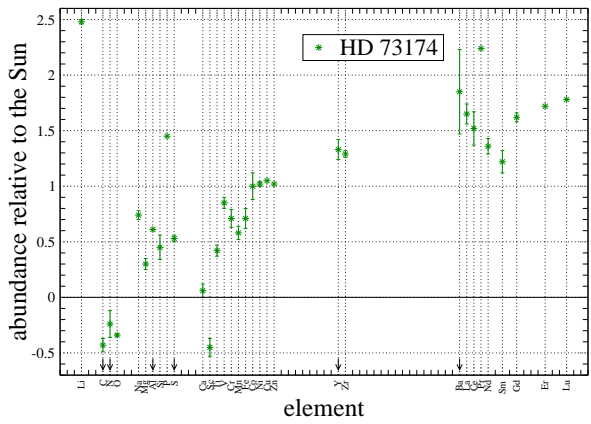}}
\caption{As for Fig.~\ref{abbondanze per hd73430} for HD~73174}
\label{abbondanze per hd73174}
\end{center}
\end{figure}

\begin{figure}[ht]
\begin{center}
\resizebox{\hsize}{!}{\includegraphics{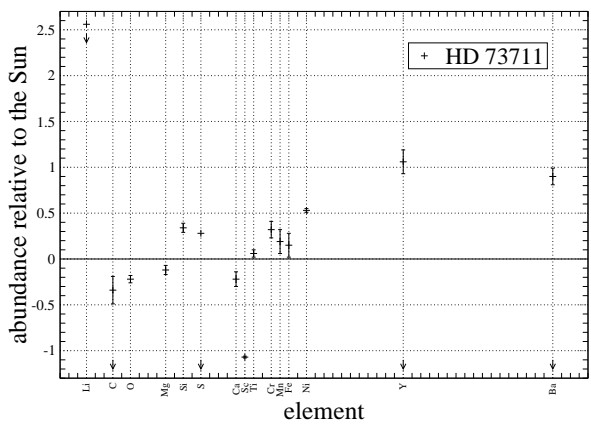}}
\caption{As for Fig.~\ref{abbondanze per hd73430} for HD~73711}
\label{abbondanze per hd73711}
\end{center}
\end{figure}

\begin{figure}[ht]
\begin{center}
\resizebox{\hsize}{!}{\includegraphics{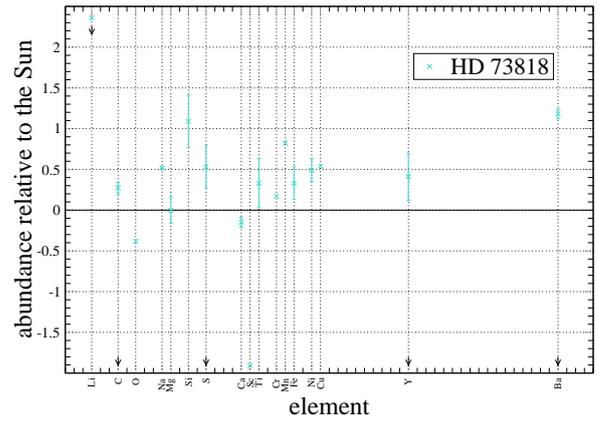}}
\caption{As for Fig.~\ref{abbondanze per hd73430} for HD~73818}
\label{abbondanze per hd73818}
\end{center}
\end{figure}

\begin{figure}[ht]
\begin{center}
\resizebox{\hsize}{!}{\includegraphics{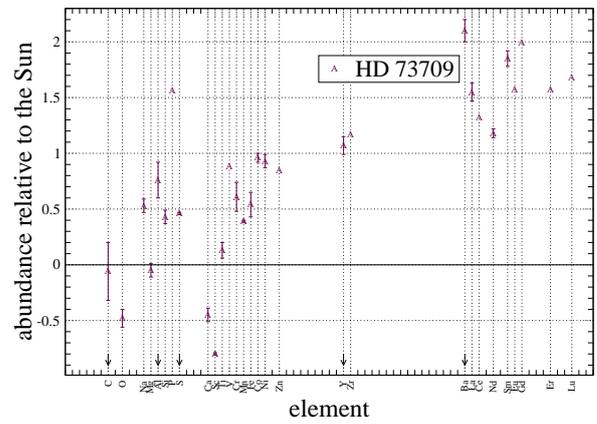}}
\caption{As for Fig.~\ref{abbondanze per hd73430} for HD~73709}
\label{abbondanze per hd73709}
\end{center}
\end{figure}

We have calculated precise Li abundances for HD~73045 and for HD~73174 taking 
into account hyperfine structure, while for HD~73730 we give an estimation
derived from one line and without hyperfine structure.
For other stars it was possible to derive only an upper limit, 
since the higher \vsini\, values did not allow a more accurate abundance 
determination. In HD~73666 we have not found any Li line, while for HD~73709 
the \mus\, spectrum did not cover the strong Li 6707.761~\AA\, line.
Our determinations confirm the overabundances (relative to the Sun) for 
three Am stars. Since for the normal A-type stars and the other Am stars we 
can only derive upper limits, we cannot confirm the overabundances.
In Fig.~\ref{livsteff} we show the absolute Li abundance as a function of 
\Teff, in comparison with the values given by \citet{burkhart1995}.
Our sample is not large enough to address the behavior of the lithium 
abundance in the temperature range between 7000\,K and 8500\,K, but we notice
that the abundances appear to be higher than those derived by 
\citet{burkhart1995} in open clusters. We cannot confirm the tendancy of the 
normal A-type stars to have a higher Li abundance. Our results appear to be 
more consistent with those derived by \citet{north2007} for field stars. 
\subsection{Comparison with previous abundance determinations}\label{comparison}
In the past years, some of the stars belonging to our sample were 
analysed by different authors: \citet{hui1997}, \citet{hui1998}, 
\citet{andri1998} and \citet{burkcoupry1998}.

In Table~\ref{comparison,table} (available only in the online version) we 
compare our final abundances with those obtained by other authors. The 
abundances coming from other papers were converted to our 
$\log{N_{{\rm el}}/N_{{\rm total}}}$ scale, using the solar abundances taken 
as reference in each paper. 

As explained in Sect.~\ref{howparameters}, our \Teff\, and \logg\, were 
determined spectroscopically, while the other authors derived these parameters
from photometry. All the measurements are in perfect agreement considering 
the associated errors. This shows that for normal A-type stars and for Am 
stars the fundamental parameters, derived from photometry, are not too far from
a more precise determination given by the spectroscopy.  
For some stars, the microturbulence velocity appears to differ for different
authors. It is notable that the our \vmic\ and that derived by 
\citet{hui1998} for the slow rotator HD~73174 are in good agreement. This
agreement is due probably to the similar methods used to derive it 
\citep[see][]{hui1998}.

Taking into account a default error of 0.2 dex for the abundances without
an individual error, inspection of Table~\ref{comparison,table} reveals a 
good agreement between our work and the previous ones. Only Li, Sc, Ba and 
Eu show some deviations. The deviation associated with the Sc abundance occurs 
because of the difficulty to calculate this abundance for Am stars in which 
Sc is underabundant, showing only shallow lines. Concerning the Ba and Eu 
discrepancies, they could be due to a different line selection with 
different \nlte\ effects associated with the Ba lines and a different 
weight given to the blended Eu lines, selected to determine the abundances. 
We are not able to find a clear reason of the discrepancies for the Li 
abundances.

We believe that our determination of the fundamental parameters and of the 
elemental abundances are affected by smaller external errors than those of the 
other authors. We state this based on the wider wavelength range available, 
on the use of better and more carefully-checked line parameters, on the more 
reliable atmospheres models used and finally on the methodology used to 
determine the abundances (fitting of the line profile instead of using 
equivalent widths).
\section{Conclusions}\label{conclusion}
We have obtained high resolution, high SNR spectra for eleven A-type stars 
belonging to the nearby intermediate-age Praesepe open cluster. For seven 
stars of the sample the spectra were also obtained in circular polarization. 

Eight stars of the sample were previously classified as Am stars, two as 
normal A-type stars and one as an Ap\,(Si) star. 
We have calculated the fundamental parameters and performed a detailed 
abundance analysis for each star of the sample. For the stars observed in 
circular polarization we have used the LSD technique to measure the 
longitudinal magnetic field. 

No significant magnetic field signature was found applying the LSD technique 
to the Am stars, which were observed with \espa\ and \mus. This leads us 
to the conclusion that peculiar abundances and slow rotation of the Praesepe 
Am stars cannot be explained by the presence of a magnetic field. But since 
nearly all of them are SBs we can conclude that slow rotation, induced by 
binarity, can lead to the Am chemical peculiarities.

Fundamental parameters and elemental abundances were derived by adjusting 
synthetic spectra to the observed ones. The model atmospheres were computed 
by means of the 8.4 version of the \llm\ code, without taking into 
account \nlte\ corrections, using atomic line data extracted from the VALD 
database. Synthetic spectra were produced with Synth3 \citep{synth3}.

The abundance analysis shows that the Blue Straggler HD~73666, previously 
classified Ap\,(Si) star, is a normal A-type star. This star does not show any
magnetic field nor peculiarity.
We confirm the membership and the classification of the two normal A-type 
stars HD~73430 and HD~73575.
HD~72942 cannot be confirmed either as member of the cluster or as an Am star.
For all the other Am stars classification is confirmed. 

For this sample of Am stars we have compared our abundances with the 
predictions of diffusion models by \citet{richer2000}. We obtained an excellent 
agreement, within the errors, with the predictions for almost all the common 
elements. Only Na and probably S show a clear discrepancy from the 
theoretical abundances. Unfortunately, our Na and S abundances do not take 
into account \nlte\ effects that would bring the abundances closer to the 
predictions of theory \citep{kamp2001}. A \nlte\ abundance determination 
should be performed, at least for a few Am stars, in order to confirm the 
abundance predictions of these two elements. \citet{richer2000} did not 
analyse Sc because it is not present in the OPAL opacity database. Model 
predictions for this element and for heavy elements are especially important 
and should be performed.

Finally, we have turned our attention to the Li abundance of the sample, 
obtaining results comparable to those of \citet{burkhart1995}, in open 
clusters, and \citet{north2007}, for field stars. 

Our results indicate that radiative diffusion, combined with turbulent mixing, 
can account for most of the chemical peculiarities found in Am stars. 
More abundance determinations on cluster stars with a variety of ages will 
help to constrain the physical processes at work in A and Am stars.
\begin{acknowledgements}
We are deeply indebted to Tanya Ryabchikova for her close and accurate
supervision of the abundance analysis. 
LF and OK have received support from the Austrian Science Foundation 
(FWF project P17980-N2). SK, JDL and GAW acknowledge support from the Natural 
Science and Engineering Council of Canada (NSERC). GAW acknowledges support 
from the Department of National Defence Academic Research Programme (DND-ARP).
This paper was based on observations obtained using the \espa\,
spectropolarimeter at the Canada-France-Hawaii Telescope (Canada), the ELODIE
spectrograph at the Observatoire de Haute Provence (France) and the \mus\, 
spectropolarimeter at the Bernard Lyot Telescope (France). 
\end{acknowledgements}
%

%

\Online
\begin{table*}[ht]
\caption[ ]{The Table shows the comparison between the abundances described in 
this paper and those derived by other authors for the common stars. We 
compare also the fundamental parameters. HBA: \citet{hui1997}; 
HA: \citet{hui1998}; A: \citet{andri1998}; BC: \citet{burkcoupry1998}; 
TW: This Work.}
\label{comparison,table}
\begin{center}
\rotatebox{0}{
\scriptsize{
\begin{tabular}{cccccccccccc}
\hline
\hline
        &\Teff&\logg&\vsini&\vmic&He	     &Li         &C          &O	         &Na	     &Mg         &Al         \\ 
\hline
HD~73666&     &     &	   &	 &	     &	         &           &           &           &           &	     \\
A       &9600 &3.80 &30    &3.0  &$-1.05(-)$ &	         &$-3.59(-)$ &$-3.37(-)$ &$-5.41(-)$ &           &	     \\
BC      &9500 & $-$ & $-$  &4.5	 &	     &$<-8.59(-)$&           &           &           &           &	     \\
TW      &9382 &3.78 &10    &1.9  &$-1.20(05)$& $-$       &$<-3.25(03)$&$-3.06(07)$&$-5.47(01)$&          &	     \\
\hline
HD~72942&     &     &	   &	 &	     &	         &           &           &           &           &	     \\
HA      &8130 &3.80 &70    &2.0  &	     &	         &           &           &           &$-4.38(-)$ &	     \\
BC      &8300 & $-$ & $-$  &4.5	 &	     &$<-8.74(-)$&           &           &           &           &	     \\
TW      &8450 &3.90 &73    &2.4  &	     &$<-8.49(-)$&           &           &           &$-4.51(03)$&	     \\
\hline
HD~73045&     &     &	   &	 &	     &	         &           &           &           &           &	     \\
HBA     &7520 &4.27 &10    &3.5  &	     &	         &           &           &           &$-4.31(-)$ &	     \\
BC      &7500 & $-$ & $-$  &4.5	 &	     &$-9.04(-)$ &           &           &           &           &$-5.44(-)$ \\
TW      &7570 &4.05 &10    &3.6  &	     &$-8.86(02)$&           &           &           &$-4.52(04)$&$<-5.41(01)$\\
\hline
HD~73730&     &     &	   &	 &	     &	         &           &           &           &           &	     \\
HA      &8020 &3.90 &32    &4.0  &	     &	         &           &           &           &$-4.67(-)$ &	     \\
BC      &8020 & $-$ & $-$  &4.5	 &	     &$-9.04(-)$ &           &           &           &           &$-5.64(-)$ \\
TW      &8070 &3.97 &29    &2.6  &	     &$-8.74(-)$ &           &           &           &$-4.43(06)$&$<-5.37(-)$\\
\hline
HD~73618&     &     &	   &	 &	     &	         &           &           &           &           &	     \\
HBA     &8060 &3.87 &60    &3.0  &	     &	         &           &           &           &$-3.93(-)$ &	     \\
A       &8050 &4.00 &60    &3.0  &	     &	         &$-3.89(-)$ &$-3.77(-)$ &$-5.71(-)$ &           &	     \\
BC      &8100 & $-$ & $-$  &4.5	 &	     &$-8.94(-)$ &           &           &           &           &$-5.29(-)$ \\
TW      &8170 &4.00 &47    &2.5  &	     &$<-8.58(-)$&$<-4.16(16)$&$-3.65(04)$&$-5.34(03)$&$-4.29(-)$& $-$	     \\
\hline
HD~73174&     &     &	   &	 &	     &	         &           &           &           &           &	     \\
HA      &8090 &4.00 &$<$10 &2.8  &	     &	         &           &           &           &$-4.40(-)$ &	     \\
BC      &8090 & $-$ & $-$  &4.5	 &	     &$-9.04(-)$ &           &           &           &           &$-5.34(-)$ \\
TW      &8350 &4.15 &$<$5  &2.9  &	     &$-8.51(02)$&           &           &           &$-4.21(05)$&$<-5.06(-)$ \\
\hline
HD~73711&     &     &	   &	 &	     &	         &           &           &           &           &	     \\
BC      &8290 & $-$ & $-$  &4.5	 &	     &$<-8.64(-)$&           &           &           &           &	     \\
TW      &8020 &3.69 &62    &2.5	 &	     &$<-8.43(-)$&           &           &           &           &	     \\
\hline
HD~73818&     &     &	   &	 &	     &	         &           &           &           &           &	     \\
BC      &7230 & $-$ & $-$  &4.5	 &	     &$-8.99(-)$ &           &           &           &           &	     \\
TW      &7230 &3.82 &66    &2.8  &	     &$<-8.63(-)$&           &           &           &           &	     \\
\hline    
HD~73709&     &     &	   &	 &	     &	         &           &           &           &           &	     \\
HBA     &8060 &3.95 &20    &3.5  &	     &	         &           &           &           &$-4.39(-)$ &	     \\
BC      &8080 & $-$ & $-$  &4.5	 &	     &$-8.64(-)$ &           &           &           &           &$-5.14(-)$ \\
TW      &8070 &3.78 &10    &2.3  &	     & $-$	 &           &           &           &$-4.56(06)$&$<-4.91(16)$\\
\hline
\hline
        &Si	    &S	        &Ca         &Sc	        &Cr	    &Mn	       &Fe         &Ni         &Ba	   &Eu	       &\\
\hline
HD~73666&	    &	        &	    & 	        &	    &	       &	   &           &	   &	       &\\
A       &$-4.89(-)$ &	        &$-5.68(-)$ & 	        &	    &	       &$-4.44(-)$ &           &$-9.31(-)$ &	       &\\
BC      &	    &$-4.64(-)$ &$-5.24(-)$ & 	        &	    &	       &$-3.94(-)$ &           &	   &$-10.04(-)$&\\
TW      &$-4.30(01)$&$<-4.47(-)$&$-5.53(14)$& 	        &	    &	       &$-4.30(07)$&           &$<-9.13(04)$& $-$       &\\
\hline
HD~72942&	    &	        &	    & 	        &	    &	       &	   &           &	   &	       &\\
HA      &	    &	        &$-6.09(-)$ &$-9.05(-)$ &$-6.40(-)$ &	       &$-4.59(07)$&$-5.09(-)$ &	   &	       &\\
BC      &	    &$-4.54(-)$ &	    & 	        &	    &	       &$-4.29(-)$ &           &	   &	       &\\
TW      &	    &$<-4.48(-)$&$-5.58(12)$&$-8.88(11)$&$-6.41(16)$&	       &$-4.47(07)$&$-5.57(-)$ &	   &	       &\\
\hline
HD~73045&	    &	        &	    & 	        &	    &	       &	   &           &	   &	       &\\
HBA     &	    &	        &$-6.22(27)$&$-9.87(-)$ &$-5.67(-)$ &	       &$-3.95(14)$&$-4.99(-)$ &	   &	       &\\
BC      &$-4.29(-)$ &$-4.94(-)$ &	    & 	        &	    &	       &$-4.17(-)$ &$-5.09(-)$ &	   &$-10.54(-)$&\\
TW      &$-4.21(05)$&$<-4.67(09)$&$-6.46(04)$&$-9.39(01)$&$-5.63(16)$&	       &$-4.03(04)$&$-4.93(06)$&	   &$-10.20(-)$&\\
\hline
HD~73730&	    &	        &	    & 	        &	    &	       &	   &           &	   &	       &\\
HA      &	    &	        &$-6.24(-)$ &$-10.54(-)$&$-6.06(-)$ &	       &$-4.35(06)$&$-5.14(-)$ &	   &	       &\\
BC      &$-4.39(-)$ &$-4.64(-)$ &	    & 	        &	    &	       &$-4.14(-)$ &$-5.34(-)$ &	   &$-10.74(-)$&\\
TW      &$-4.21(05)$&$<-4.54(03)$&$-6.36(03)$&$-9.71(03)$&$-5.65(08)$&	       &$-4.14(05)$&$-5.14(07)$&	   &$-10.42(-)$&\\
\hline
HD~73618&	    &	        &	    & 	        &	    &	       &	   &           &	   &	       &\\
HBA     &	    &	        &$-5.70(-)$ &$-10.53(-)$&$-5.92(-)$ &	       &$-4.08(20)$&$-4.93(-)$ &	   &	       &\\
A       &$-4.69(-)$ &$-4.51(-)$ &$-5.78(-)$ & 	        &	    &$-6.75(-)$&$-4.24(-)$ &$-5.49(-)$ &$-9.21(-)$ &	       &\\
BC      &$-4.14(-)$ &$-4.74(-)$ &	    & 	        &	    &	       &$-4.01(-)$ &           &	   &	       &\\
TW      &$-4.51(-)$ &$<-4.52(13)$&$-5.74(07)$&$-9.51(27)$&$-5.93(10)$&$-6.28(-)$&$-4.13(10)$&$-5.10(15)$&$<-8.09(18)$&	       &\\
\hline
HD~73174&	    &	        &	    & 	        &	    &	       &	   &           &	   &	       &\\
HA      &	    &	        &$-6.07(07)$&$-10.52(-)$&$-6.09(-)$ &	       &$-4.27(16)$&$-5.20(-)$ &	   &	       &\\
BC      &$-4.34(-)$ &$-4.74(-)$ &	    & 	        &	    &	       &$-4.22(-)$ &$-5.24(-)$ &	   &$-10.54(-)$&\\
TW      &$-4.08(11)$&$<-4.37(03)$&$-5.70(08)$&$-9.44(08)$&$-5.69(08)$&	       &$-3.88(09)$&$-4.79(02)$&	   & $-$       &\\
\hline
HD~73711&	    &	        &	    & 	        &	    &	       &	   &           &	   &	       &\\
BC      &	    &$-4.64(-)$ &	    & 	        &	    &	       &$-3.94(-)$ &           &	   &	       &\\
TW      &	    &$<-4.62(-)$ &	    & 	        &	    &	       &$-4.44(13)$&           &	   &	       &\\
\hline
HD~73818&	    &	        &	    & 	        &	    &	       &	   &           &	   &	       &\\
BC      &	    &	        &	    & 	        &	    &	       &$-4.34(-)$ &           &	   &	       &\\
TW      &	    &	        &	    & 	        &	    &	       &$-4.26(20)$&           &	   &	       &\\
\hline  
HD~73709&	    &	        &	    & 	        &	    &	       &	   &           &	   &	       &\\
HBA     &	    &	        &$-5.92(17)$&$-10.20(-)$&$-5.72(-)$ &	       &$-3.96(17)$&$-4.76(-)$ &	   &	       &\\
BC      &$-4.24(-)$ &$-4.54(-)$ &	    & 	        &	    &	       &$-4.04(-)$ &$-5.04(-)$ &	   &$-10.24(-)$&\\
TW      &$-4.10(06)$&$<-4.44(01)$&$-6.18(06)$&$-9.79(01)$&$-5.79(13)$&	       &$-4.05(11)$&$-4.88(06)$&	   &$-9.95(-)$ &\\
\hline
\end{tabular}
}
}
\end{center}
\end{table*}


\end{document}